\documentclass[epj]{svjour}
\usepackage{amsmath,amssymb,amsbsy}
\usepackage[applemac]{inputenc}
\usepackage{graphics}
\usepackage{epsfig}
\journalname{The European Physical Journal E}
\newcommand{\bea}{\begin{eqnarray}}
\newcommand{\eea}{\end{eqnarray}}
\newcommand{\bc}{\begin{center}}
\newcommand{\ec}{\end{center}}
\newcommand{\btab}{\begin{tabular}}
\newcommand{\etab}{\end{tabular}}

\let\oldepsilon\epsilon
\let\epsilon\varepsilon
\let\varepsilon\oldepsilon

\begin{document}
\title{On dense granular flows.}
\author{G.D.R Midi}
\institute{Groupement De Recherche Milieux Divis\'es, CNRS, GDR2181,France, \\
gdrmidi@polytech.univ-mrs.fr}
\date{\today}

\abstract{The behaviour of dense assemblies of dry grains submitted to continuous shear deformation has been the subject of many experiments and discrete particle simulations. This paper is a collective work carried out among the French research group GDR Milieux Divis\'es. It proceeds from the collection of results on steady uniform granular flows obtained by different groups in six different geometries both in experiments and numerical works. The goal is to achieve a coherent presentation of the relevant quantities to be measured i.e. flowing thresholds, kinematic profiles, effective friction, etc. First, a quantitative comparison between data coming from different experiments in the same geometry enforces the robust features in each case. Second, a transversal analysis of the data across the different configurations, allows us to identify the relevant dimensionless parameters, the different flow regimes and to propose simple interpretations. The present work, more than a simple juxtaposition of results, underlines the richness of granular flows and enhances the open problem of defining a single rheology.}
 \authorrunning{G.D.R. Milieux Divis\'es} \titlerunning{On dense granular flows.} \maketitle

\section{Introduction}

At the frontier between physics and mechanics, the flow of granular materials has become a very active research domain \cite{BR70,D99,HHL98,CR00,dG98,THNS82,S89,HR94,R00}. The behaviour of assemblies of grains can be very complex even in the simple case of dry cohesionless particles. When the grains are large enough ($d>250\mu m$) and the surrounding fluid is not too viscous, the particle interactions are dominated by contact interactions. Capillary forces, van der Waals forces or viscous interactions can be neglected and the mechanical properties of the material are only controlled by the momentum transfer during collision or frictional contacts between grains.

Still, the flows of these dry granular materials are not easy to describe. They are usually divided in three classes depending on the flow velocity. First a quasi-static regime where grain inertia is negligible. The material is often described using soil plasticity models \cite{N92,SW68}. Secondly, a "gaseous" regime exists when the medium is strongly agitated and the grains far apart one from another. In this regime particles interact through binary collisions and a kinetic theory has been developed by analogy with the kinetic theory of gases \cite{C90,G99}. In between these two regimes there exists a dense flow regime where grain inertia becomes important but where a contact network still exists that percolates through particles \cite{PC02}. Up to now no constitutive equations are available in this "liquid" regime and no unified framework allows to describe the whole dynamics from quasi-static to gaseous regime.

The lack of information about the liquid regime and about the transition between the different regimes has recently motivated many experimental, numerical and theoretical works. Different flow configurations have been investigated from confined flows in channels to free surface flows on piles, both experimentally and numerically. However, although important and precise information is now available about the flow characteristics, it is often difficult to extract common features and general trends for granular flows. Configurations are not the same, experimental or numerical conditions varies from one study to another. In this paper we take advantage of a French research network supported by the CNRS, the GDR Milieux Divisés, to collect the data from different groups.

First, we plan to compare the data obtained under different experimental or numerical conditions, in order to extract the most robust features. What are the relevant flow characteristics, i.e. thresholds, kinematic profiles, effective friction, etc, in the different flow configurations? How do these quantities depend on the details of the experimental set up or numerical procedures? Second, we would like to sort the different flow configurations according to the common features and differences that arise among them. What are the relevant time and length scales in the different configurations? Are there underlying common physical phenomena controlling flow properties in the different geometries? As a result, we expect to identify simple and basic features that could help in developing future model for dense granular flows.

Let us emphasise that this collective work does not pretend to be exhaustive. First, the paper focus only on steady uniform flows of slightly polydispersed grains, leaving aside very important questions such as avalanche triggering, intermittent flows or segregation.
Second, since the data presented here come from the research group GDR Midi and collaborators, many important contributions are not included. We refer to them in the citations. However, the huge activity in the domain makes the exercise difficult. We take refuge behind this excuse for all the contributions that have been omitted.

\section{Six different configurations}

\begin{figure}[h!] \bc
\includegraphics{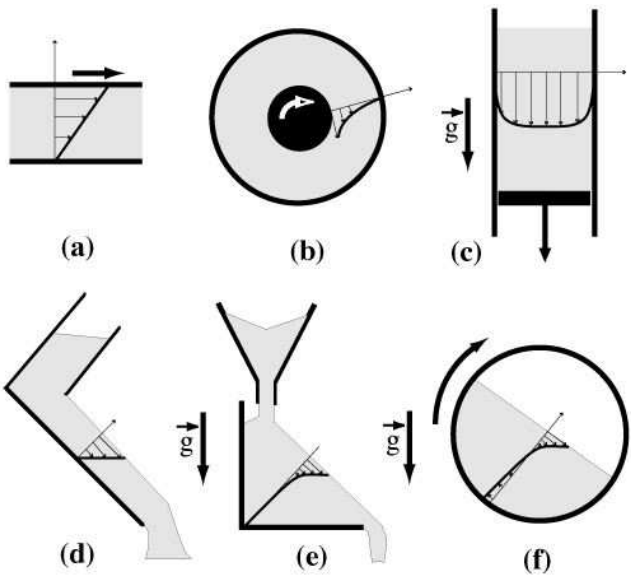}
\caption{The six configurations of granular flows: (a) plane shear, (b) annular shear, (c) vertical chute flows, (d) Inclined plane, (e) heap flow, (f) rotating drum.}
\label{figintro}
\ec
\end{figure}

Dense granular flows are mainly studied in six different configurations (Fig.~\ref{figintro}), where a simple shear is achieved and rheological properties can be measured. These geometries are divided in two families: confined and free surface flows.

The confined flows are the plane shear geometry (Fig.~\ref{figintro}a) where a shear is applied due to the motion of one wall, the annular shear (Fig.~\ref{figintro}b) where the material confined in between two cylinders is sheared by the rotation of the inner cylinder and the vertical chute flow configuration (Fig.~\ref{figintro}c) where material flows due to the gravity in between two vertical rough walls. Free surface flows are flow of granular material on a rough inclined plane (Fig.~\ref{figintro}d), flow at the surface of a pile (Fig.~\ref{figintro}e) and flow in a rotating drum (Fig.~\ref{figintro}f). The driving force is in these last three cases the gravity.
In the following we consider successively the six configurations. The data comes from different experiments and numerical simulations briefly described in a table at the beginning of each section. We report for each of them the flowing threshold, the kinematic properties (velocity $V(y)$, volume fraction $\Phi(y)$ and velocity fluctuations $\delta V^2(y)$ profiles) and the rheological behaviour, before discussing the influence of the various experimental or numerical parameters. Both the notations and the dimensionless quantities naturally used to present the results are given in appendix A.

\section{Plane shear flow}
\begin{figure*}[ht!] \bc \vspace{10 mm}
\includegraphics{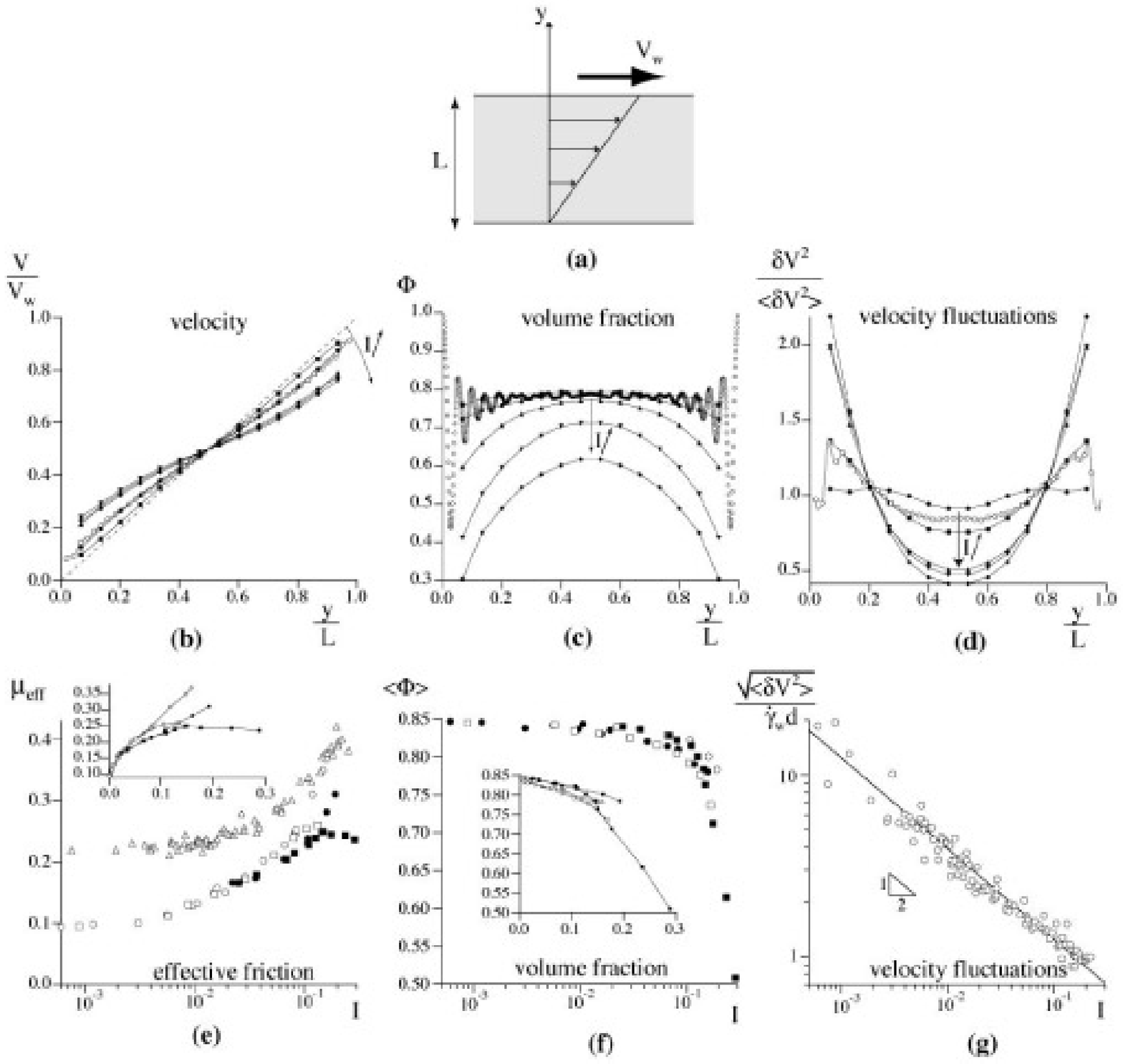}
\caption{
{\bf Plane shear.}
{\bf (a)} Setup.
{\bf (b)} Velocity profiles rescaled by the wall velocity from simulations
PS1: $I$ between $0.2$ and $0.5$ ($e=0.8$) (black symbols) and
PS2: $I=0.23$ ($e=0.1$, $\mu_{\rm p}=0.4$) ({\Large $\circ$}).
{\bf (c)} Volume fraction profiles as a function of $I$ from simulations
PS1: $I$ between $0.2$ and $0.5$ ($e=0.8$) (black symbols) and
PS2: $I=0.23$ ($e=0.1$, $\mu_{\rm p}=0.4$) ({\Large $\circ$}).
{\bf (d)} Velocity fluctuations profile normalised by its mean
value across the cell as a function of $I$ from simulations
PS1: $I$ between $0.06$ and $0.5$ ($e$ between $0.6$ and $0.98$) (black symbols),
 and
PS2: $I=23$ ($e=0.1$, $\mu_{\rm p}=0.4$) ({\Large $\circ$}).
{\bf (e)} Effective friction $\mu_{\rm eff}$ as a function of $I$ measured at the wall in simulation
PS1: $e=0.1$ ({\Large $\bullet$}) and $e=0.9$ ($\blacksquare$) and inside the flow in simulation
PS2: $e=0.1$ ({\Large $\circ$}), $e=0.9$ ($\square$), both for
$\mu_{\rm p}=0$. ($\triangle$) correspond to simulations PS2 with $0.1 <\mu_{\rm p} < 0.8$. Inset: same
curves in linear-linear representation.
{\bf (f)} Mean volume fraction $<\phi>$ as a function of $I$ from simulations
PS1: $e=0.1$ ({\Large $\bullet$}) and $0.8$ ($\blacksquare$) and
PS2: $e=0.1$ ({\Large $\circ$}) and $0.9$ ($\square$). Inset: same
curves in linear-linear representation.
{\bf (g)} Relative velocity fluctuations as a function of $I$ from simulations
PS2: $e=0.1$ and $0.9$, $\mu_{\rm p}=0$, $0.4$ and $0.8$.
}
\label{FigPlaneShear}
\ec
\end{figure*}

\subsection{Set-up}
In the aim of studying flow rheology, the plane shear (Fig.~\ref{FigPlaneShear}a) is conceptually the simplest geometry one naturally thinks of. The flow is obtained between two parallel rough walls, a distance $L$ apart and moving at the relative velocity $V_{\rm w}$. In the following we note $\dot\gamma_{\rm w}=V_{\rm w}/L$ the mean shear rate. In this configuration, the stress distribution is uniform inside the sheared layer. However, because of gravity, this homogeneous state is not achieved in existing experiments~\cite{SS84,HI85} but is obtained in discrete particles simulations. Most of the results found in the literature are obtained imposing the wall velocity and measuring the shear stress~\cite{BSS90,TG91,ZC92,SHS98,AS02}. Some are carried out controlling the shear force applied to the moving wall in order to study the flow thresholds~\cite{VTA03}.

In the following, we present results of two dimensional discrete particles simulations where $V_{\rm w}$ is imposed and the number of grains (size $d$ and mass $m$) within the cell is fixed (periodic boundary conditions are used along the shear direction). The data are summarised in table~\ref{TabPS}. In one case the volume - the cell width $L$ -- and thereby the density $\rho$ -- or the volume fraction $\Phi$ -- are controlled and the pressure $P$ is measured, while in the other case the pressure is controlled and the density is measured. Once the inter-particle contact laws are fixed, the simulations depend on two parameters : the wall velocity $V_{\rm w}$ and the normal stress $P$ or the density $\rho$. This define a single dimensionless numbers describing the relative importance of inertia and confining stresses,
\begin{equation}
I=\frac{\dot\gamma_{\rm w} d}{\sqrt{P/\rho}}.
\end{equation}
Both simulations are performed in the limit of rigid grains, so that the macroscopic time scale $L/V_{\rm w}$ is much larger than the microscopic timescales i.e. the elastic and the dissipative ones. The inter-particle friction coefficient $\mu_{\rm p}$ is null when not specified. The roughness of the walls is made of glued grains similar to the flowing grains.
\begin{table*}[ht!]
\bc
\begin{tabular}{|c|c|c|c|c|c|}
\hline
$\#$ & Exp/Num & 2D/3D & Material & Boundary Conditions & Ref \\ \hline
PS1 & Num (MD) & 2D & polydisperse spheres ($\pm 10 \%$) $\mu_{\rm p}= 0$ & fixed volume ($L=15$)& \cite{IK02,DCRI03} \\ \hline
PS2 & Num (MD) & 2D & polydisperse disks ($\pm 20 \%$) & fixed pressure ($L=20 \rightarrow 100$) & \cite{DCRI03,DERC03,D04}\\ \hline
\end{tabular}
\caption{Data sources for plane shear flows. MD is for molecular dynamics.}
\label{TabPS}
\ec
\end{table*}

\subsection{Kinematic properties}

\subsubsection{Velocity profiles}
Fig.~\ref{FigPlaneShear}b displays the velocity profiles obtained in different flow regimes. As long as $I$ remains small (smaller than say $0.1$), the velocity profile $v(y)$ is linear. Accordingly the shear rate is uniform and imposed by the geometry:
\begin{equation}
\dot\gamma=\dot\gamma_{\rm w}=\frac{V_{\rm w}}{L}
\end{equation}
\noindent
For larger $I$, a slip velocity appears at the boundaries and the profile becomes slightly S-shaped.

\subsubsection{Volume fraction profile}

The above two regimes correspond to a dense flow regime at small $I$ and a collisional dilute regime at larger $I$ as shown in figure~\ref{FigPlaneShear}c. The volume fraction profiles $\Phi(y)$ are plotted for different $I$. For small $I$, $\Phi$ is uniform across the shear cell -- apart from oscillations due to ordering close to the wall. By contrast, for larger $I$, a significant decrease of $\Phi$ is observed close to the walls so that it is no longer uniform.

The transition is evidenced on figure~\ref{FigPlaneShear}f, where the average volume fraction is plotted as a function of $I$. $<\Phi>$ reaches its maximum, $0.85$, in the quasi-static limit $I\simeq 0$, decreases gently linearly with $I$ down to $0.80$ for $I\simeq 0.1$ and decreases more rapidly for larger $I$. Small $I$ corresponds to the dense flow regime, associated to a network of enduring contacts~\cite{RC02}, and large $I$ corresponds to the dynamic inertial regime, associated to binary collisions~\cite{IK02}.

\subsubsection{Velocity fluctuations profile}

The velocity fluctuations profiles are shown in figure~\ref{FigPlaneShear}d, where $\delta V^2$ is normalised by its mean value across the shear cell. Here again the profiles are uniform for small $I$, whereas a significant increase of the fluctuations is observed close to the walls for larger $I$.

Figure~\ref{FigPlaneShear}g shows that the root mean square (rms) velocity exhibits an interesting scaling with both shear rate and pressure. The following scaling law is observed:
\begin {equation}
<\delta V^2> \propto d{\dot\gamma_{\rm w}} \sqrt{\frac{P}{\rho}}.
\end{equation}
The velocity fluctuations depend on both the shear rate and the confining pressure. This means that the relative velocity fluctuations depends on the sole dimensionless number $I$:
\begin{equation}
\frac {\sqrt{<\delta V^2>}}{\dot\gamma_{\rm w} d} \propto I^{-1/2}
\end{equation}
Let us underline that the above scaling for the mean value of the velocity fluctuations is valid up to the largest value of $I$, although in this regime the velocity profiles are not uniform across the channel.

\subsection{Effective friction}
The effective friction coefficient $\mu_{\rm eff}$ is defined either as the ratio of shear to normal force at the wall, or as the ratio of the shear stress to the pressure inside the material. Both definitions give approximately the same results. Figure~\ref{FigPlaneShear}e displays the effective friction for two different simulations and different values of the restitution coefficient $e$ and interparticle friction $\mu_{\rm p}$. It shows that $\mu_{\rm eff}$ starts from a finite value $\mu_S$, corresponding to the internal Mohr-Coulomb friction~\cite{N92}, remains approximately shear rate independent in the quasi-static regime ($I<10^{-3}$), and increases for larger values of $I$~\cite{BSS90,TG91,TKS98,AS02} up to some threshold where the flow leaves the dense regime. Above this threshold $\mu_{\rm eff}$ saturates or even slightly decreases. This threshold value depends on the restitution coefficient $e$ as observed in inset of figure~\ref{FigPlaneShear}e. Whereas for $e=0.9$ the transition occurs for $I\simeq 0.1$, for $e=0.1$ the dense flow regime extends up to the maximum value of $I$ explored in the simulation.

\subsection{Parametric study}
We now discuss the influence of the microscopic coefficients, namely the restitution coefficient $e$ and the interparticle friction $\mu_{\rm p}$. The major result is that in the dense flow regime, the volume fraction, the velocity profiles and the effective friction neither depend on $e$ nor $\mu_{\rm p}$ as long as $\mu_{\rm p}$ is of order $1$ (say larger than $0.1$). If $\mu_{\rm p}=0$ one simply observes (see Fig.~\ref{FigPlaneShear}e) a shift of the effective friction towards lower values~\cite{DERC03,D04}. However, as already mentioned, the transition from the dense flow regime to the dilute collisional one depends on the restitution coefficient $e$~\cite{IK02}. This is clearly observed on the effective friction dependence on $I$ (inset of Fig.~\ref{FigPlaneShear}e) as well as on the volume fraction dependence on $I$ (inset of Fig.~\ref{FigPlaneShear}f).

As a conclusion, for usual granular materials ($e$ not too close to $1$ and $\mu_{\rm p}$ not to small), the flow properties (velocity, dilatancy and effective friction) in the dense regime are controlled by the dimensionless number $I$ only.

\section{Annular shear flow}
%
\begin{figure*}[ht!] \bc \vspace{5 mm}
\includegraphics{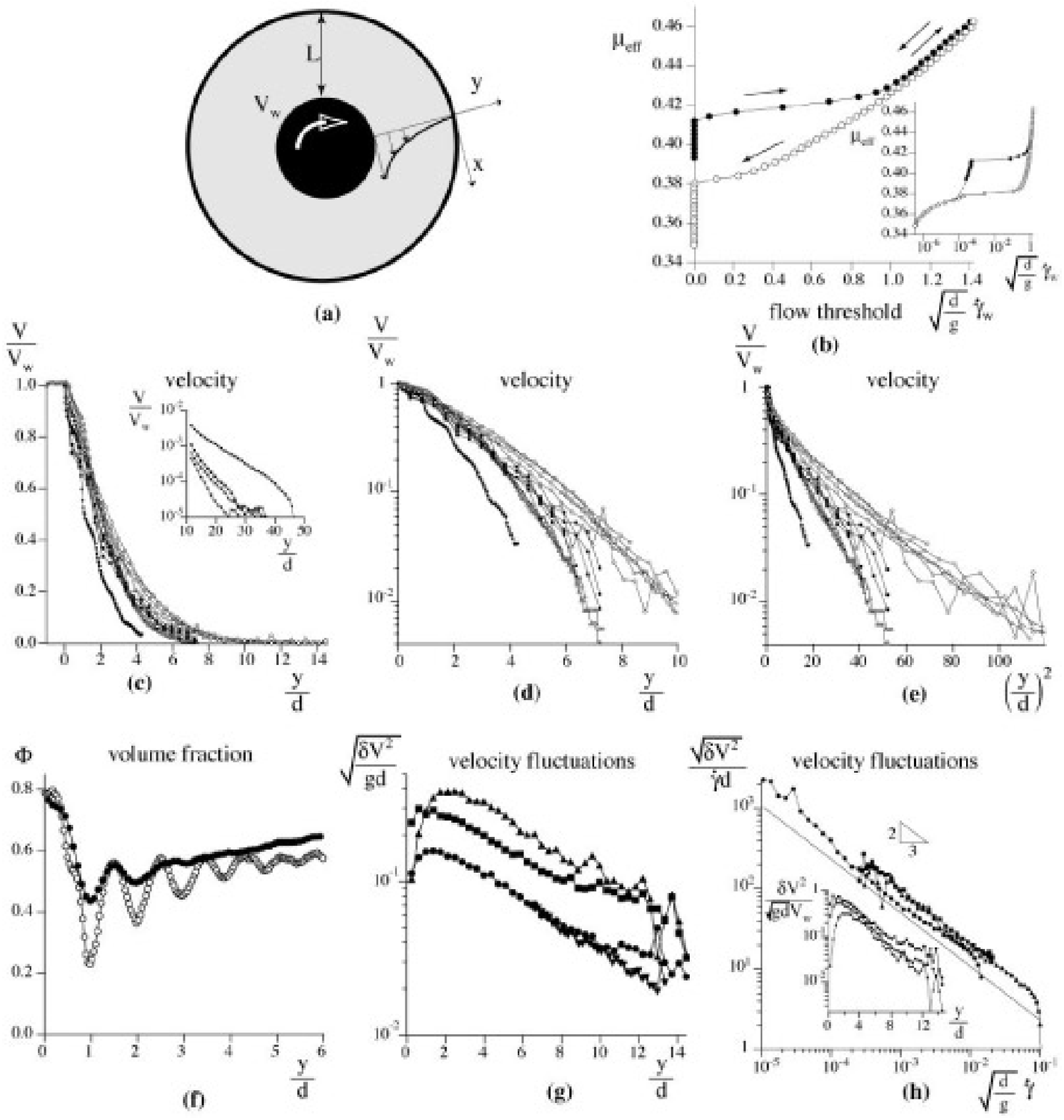}
\caption{
{\bf Annular shear.}
{\bf (a)} Set up.
{\bf (b)} Effective friction as a function of the shear rate 
from experiment AS1. Up stress ramp ({\Large $\bullet$}) and down stress ramp ({\Large $\circ$}).
Linear and Logarithmic scale (inset).
{\bf (c)} Velocity profiles, for various shear rate 
from experiments AS2: $8 \rightarrow 63 s^{-1}$ ($\blacksquare$), AS3: $25 \rightarrow 50 s^{-1}$ ($\triangle$ for poppy seeds and $\blacktriangle$ for mustard seeds), AS4: $1,7 \rightarrow 171 s^{-1}$ ({\Large $\circ$}); inset: velocity profiles from experiment AS5 for increasing displacement of the inner wall: $12-29$ mm , $29-46$ mm, $46-62$ mm , $62-79$ mm.
{\bf (d)} Same profiles and same symbols: $Log(V/V_{\rm w})$ as a function of $y/d$.
{\bf (e)} Same profiles: $Log(V/V_{\rm w})$ as a function of $(y/d)^2$.
{\bf (f)} Volume fraction profiles from experiment AS3: mustard ({\Large $\circ$}) and poppy ({\Large $\bullet$}).
{\bf (g)} Velocity fluctuations profiles from experiment AS4 for $\dot \gamma=1.7$ ({\Large $\bullet$} and $\blacktriangle$), $42$ ($\blacksquare$), $171$ s$^{-1}$ ($\triangle$). 
{\bf(h)} Same data : fluctuation rate versus dimensionless shear rate. Inset: Profiles of the rescaled fluctuations. 
}
\label{FigAnnularShear}
\ec
\end{figure*}

\subsection{Set-up}

The annular shear cell is the classical geometry used for studying rheological properties of complex fluids. In the case of granular materials, it has been extensively studied both experimentally~\cite{TKS98,MOB96,HBV99,VHB99,MDKENJ00,BLSLG02} and in discrete particle simulations~\cite{LLH00,S99}.
\begin{table*}[ht!]
\bc
\begin{tabular}{|c|c|c|c|c|c|c|c|}
\hline
$\#$ & Exp/Num & 2D/3D & Bulk particles & $R_i/d$ & $L/d$ & $W/d$ & Ref \\ \hline
AS1 & Exp & 3D & polystyrene beads & 50 & 22 & 180 & \cite{DCBC02,D04} \\ & (Rheometry) & & (d=0.25 mm) & & & & \\ \hline
AS2 & Exp & 3D & mustard seeds & 15 & 17,5 & 80 & \cite{D04} \\ & (MRI - hh) & & (d=2 mm) & & & & \\ \hline
AS3 & Exp & 3D & mustard - poppy seeds & 14 - 32 & 9 - 19 & 33 - 75 & \cite{MDKENJ00} \\ & (MRI, XT - hh) & & (d=1.8 mm - 0.8 mm) & & & & \\ \hline
AS4 & Exp & 3D & glass beads & 68 & 16 & NA & \cite{BLSLG02} \\ & (HSI - fs) & & (d=0.75 mm) & & & & \\ \hline
AS5 & Exp & 3D & sand & 100 & 100 & 100 & \cite{CSCVR03} \\ & (I - bs) & & (d=1 mm) & & & & \\ \hline
\end{tabular}
\caption{Data sources for annular shear flows. (MRI) is for Magnetic Resonance Imaging, (XT) for X-ray tomography, (HSI) for high speed imaging, (I) for imaging. In experiment AS4, the granular material is sometimes fluidised.}
\label{TabAS}
\ec
\end{table*}

In this geometry (Fig.~\ref{FigAnnularShear}a), a layer of height $W$ of granular material is sheared between two coaxial rough cylinders, with a distance $L$ (Taylor-Couette cell). The roughness of the walls is made of glued grains, similar to the flowing grains. The outer cylinder (radius $R_o$) is fixed. The inner cylinder (radius $R_i$) is moving at a rotation rate $\Omega$ so that the velocity at the inner wall is $V_{\rm w}=\Omega R_i$. As we will see below, in this geometry the shear is localised on a few particle layers close to the inner wall. Accordingly, we choose $\dot\gamma_{\rm w}=V_{\rm w}/d$ as the characteristic shear rate.

The gravity acts in the transverse direction $z$. There is usually a free surface, but the material may also be confined vertically, so that it becomes possible to control the pressure \cite{CSCVR03}. The stress distribution is characterised by an hydrostatic pressure gradient along the $z$ direction: due to the shear, the wall friction is mobilised perpendicularly to the gravity so that the Janssen effect is inactive~\cite{TKS98,D04}. In the $xy$ plane, the normal stress $P$ is uniform given that centrifugal effects are negligible. The shear stress $\tau$ decreases as $1/(y+R_i)^2$.

In this section, we present results from five experiments, whose characteristics are summarised in Tab.~\ref{TabAS}. Two kinds of experiments are performed depending whether the motion of the inner cylinder is controlled by imposing the torque $\Gamma$ or the rotation rate $\Omega$.

\subsection{Flow thresholds}

The flow thresholds are measured by first increasing, then reducing the torque applied to the inner cylinder. Figure~\ref{FigAnnularShear}b displays the effective friction $\mu_{\rm eff}$ versus the dimensionless characteristic shear rate $\dot\gamma_{\rm w} \sqrt{d/g}$. The effective friction is obtained from the torque measurements assuming an hydrostatic pressure distribution : 
\begin{equation}
\mu_{\rm eff}= \frac{\tau_{\rm w}}{P_{\rm w}} \quad {\rm with} \quad \tau_{\rm w}=\frac{\Gamma}{2\pi R_i^2 W} \quad {\rm and} \quad P_{\rm w}=\frac{1}{2} \rho g W
\end{equation}
 After a strong pre-shear, a stress ramp is applied starting from a solid state. Small stick-slip motions are observed before the flow starts at a critical torque, with a sudden jump of the rotation velocity. Above this critical torque, continuous steady flows are observed~\cite{TG91,NKBG98} and $\dot\gamma_{\rm w}$ increases with $\tau_{\rm w}$. When slowly decreasing the torque, the stress-strain relation is first reversible. Further decreasing the torque, the flow is sustained down to a lower critical stress where the flow abruptly stops. As a result, the flowing transition is strongly hysteretic~\cite{DCBC02,D04}.

\subsection{Kinematic properties}

In the experiments reported here, the flow structure has been investigated. The different profiles have been measured either at the free surface ("fs") or at the bottom surface ("bs") through a glass window, or well inside the material, at half height ("hh"), using sophisticated techniques (magnetic resonance imaging or X ray tomography) (see table \ref{TabAS}).

\subsubsection{Velocity profiles}

Figure~\ref{FigAnnularShear}c gathers measurements of velocity profiles in three experiments with different gaps $L/d$ and different shear rates. The profiles are qualitatively similar in all experiments. The shear is localised near the moving wall, and the width of the shear layer is of the order of five grains. Layering in the first layers is apparent for round grains. In each experiment, the shape of the velocity profile does not depend on the shear rate $\dot\gamma_{\rm w}$.

Both exponential and gaussian fits have been proposed for the velocity profiles~\cite{VHB99,MDKENJ00,BLSLG02}. Figure~\ref{FigAnnularShear}d shows that the velocity decays slightly faster than exponential. Also the logarithmic plot of the velocity profile versus $(y/d)^2$ (Fig.~\ref{FigAnnularShear}e) shows that the velocity profile is rather gaussian when not too close to the wall. However, the slopes and thus the shear band characteristic sizes are very scattered among the different experiments. Finally let us mention that the velocity profiles measured far from the wall during the transient establishment of the flow (see inset of Fig.~\ref{FigAnnularShear} c) recover an exponential tail. These measurements also show that the shear is localised closer and closer to the wall while the flow establishes itself.

\subsubsection{Volume fraction profiles}

As shown in figure~\ref{FigAnnularShear}f, the volume fraction slightly increases with the distance to the inner wall. Also the layering of the material close to the inner wall is more important for rounder particles (mustard seeds compared to poppy seeds).

\subsubsection{Velocity fluctuations profiles}

As shown in figure~\ref{FigAnnularShear}g the velocity fluctuations decrease exponentially with the distance to the moving wall on a typical length scale larger than the size of the shear band. This characteristic length remains constant when varying the wall shear rate. By contrast, the typical fluctuations level is shifted upwards with the wall shear rate. The typical velocity fluctuations $\delta V^2(y)$ do not scale simply with $V_{\rm w}^2$, but rather with the wall velocity as shown in inset of figure~\ref{FigAnnularShear}h. The shift between the profiles is indeed reduced when $\delta V^2(y)$ is rescaled with $V_{\rm w} \sqrt{gd}$.

In order to relate the local velocity fluctuations to the local shear rate, figure~\ref{FigAnnularShear}h displays the relative fluctuations level $\sqrt{\delta V^2(y)}/\dot\gamma d$ measured at the free surface as a function of the dimensionless shear rate $\dot\gamma(y) \sqrt{d/g}$. The data are compatible with a local relationship between these two quantities. The exponent \cite{BLSLG02} has been recovered numerically in \cite{D04}.

\section{Vertical chute flow}
\begin{figure*} \bc \vspace{5 mm}
\includegraphics{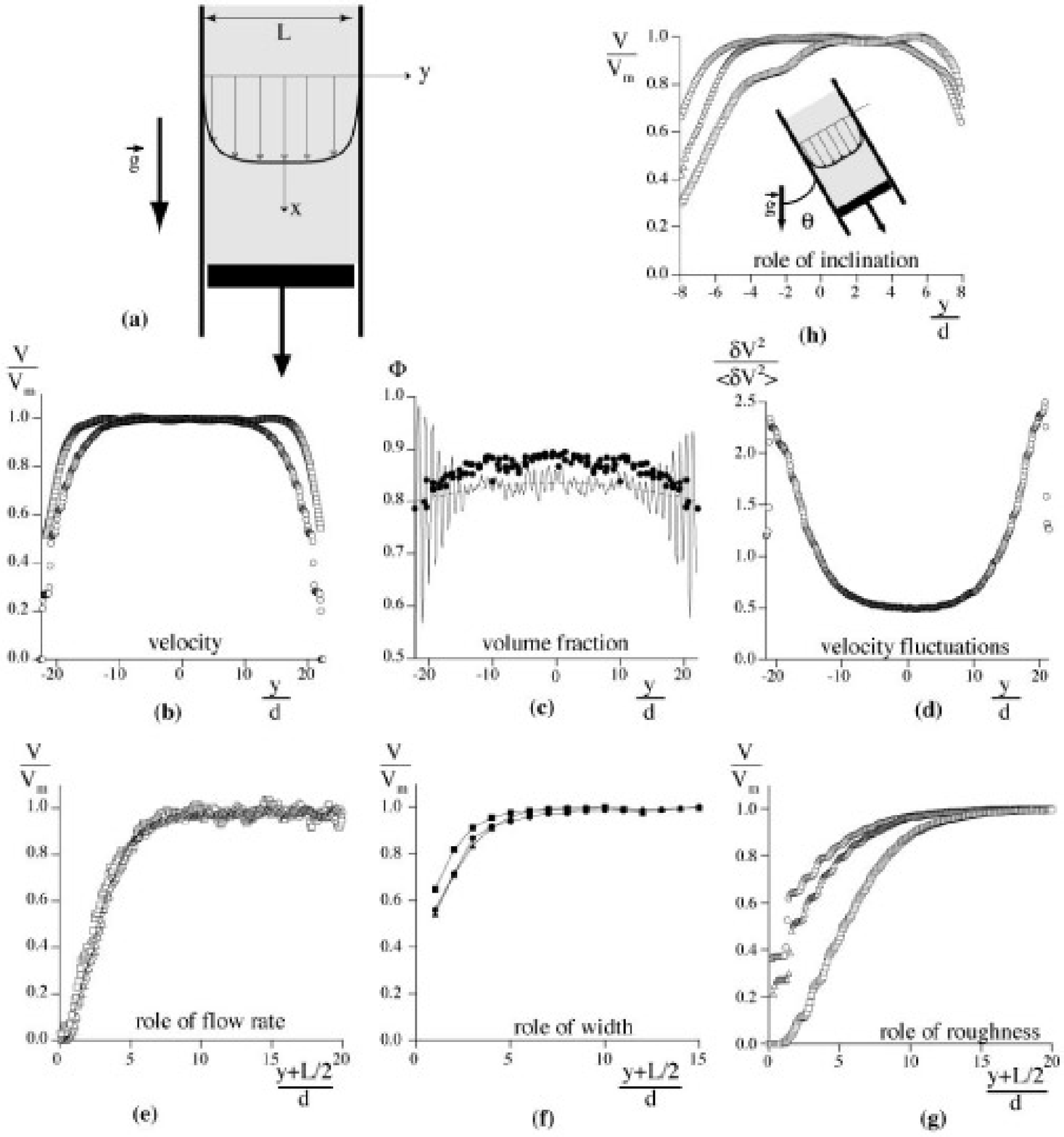}
\caption{
{\bf Vertical chute flow.}
{\bf (a)} Set up.
{\bf (b)} Typical velocity profiles for $L=45d$ from experiments VC1
({\Large $\circ$}) and from 2D simulations VC2, $d_{\rm w}=d$ ($\square$). The velocity
$V$ is rescaled by its value $V_m$ at the centre.
{\bf (c)} Volume fraction profile for $L=45d$ from experiments VC1 ({\Large $\bullet$}) and from simulations VC2, $d_{\rm w}=d$ (solid line). Dashed line
is the mean profile obtained by averaging over one particle diameter.
{\bf (d)} Typical velocity fluctuation profile normalised by its averaged
value across the channel, From VC2, $d_{\rm w}=d$.
{\bf (e)} Rescaled velocity profiles for different flow rates from VC3:
 $Vmax=11\sqrt{gd}$ ({\Large $\circ$}) , $Vmax=20\sqrt{gd}$ ($\square$),
$Vmax=30\sqrt{gd}$ ($\triangle$).
{\bf (f)} Rescaled velocity profiles for different channel width from VC1:
$L=16d$ ({\Large $\bullet$}), $L=28d$ ($\blacksquare$), $L=45d$ ($\blacktriangle$).
{\bf (g)} Rescaled velocity profiles for different wall roughness from VC3:
 $d_{\rm w}=0.5d$ ({\Large $\circ$}) , $d_{\rm w}=d$ ($\triangle$), $d_{\rm w}=4d$ ($\square$).
{\bf (h)} Rescaled velocity profiles in a channel inclined at $\theta$ from
vertical, from VC1: $\theta=0^o$ ({\Large $\circ$}) , $\theta=33^o$ ($\triangle$),
$\theta=59^o$ ($\square$).
}
\label{FigChuteFlow}
\ec
\end{figure*}

\subsection{Experimental Set-up}

Flows in silo have been extensively investigated motivated by their numerous practical applications \cite{NL80,NHT95,PG96,MD97,CPMDBGCR01,P02,DL99,BBFJ89,BGRHHH02}. In its most simplified geometry, the container reduces in three dimensions to a cylinder of diameter $L$ and in two dimensions to two parallel walls separated by a distance $L$ (Fig.~\ref{FigChuteFlow}a). Gravity drives the material down between the walls. Far from the free surface and from the bottom, the flow is uniform along the $x$ direction. The flow rate $Q$ can be controlled either by an aperture at the bottom of the device, whose opening is precisely controlled, or by moving the bottom retaining wall at a controlled velocity. The walls are made rough by gluing particles at the walls.

In this geometry the stress distribution is given by the equilibrium. If the column is high enough, Janssen effect imposes that stresses are independent of the $x$ position. Under this assumption, the normal stress $\sigma_{xx}$ is a constant whereas the tangential stress varies linearly with the distance to the walls: $\sigma_{xy}=2\tau_{\rm w} y/L$ where $\tau_{\rm w}$ is the shear stress at the wall. No shear stress exists along the symmetry axis $y=0$. Data used in this section are summarised in table~\ref{tablechute}.
\begin{table*}[ht!]
\bc
\begin{tabular}{|c|c|c|p{3.5cm}|c|p{2.5cm}|c|}
\hline
$\#$ & Exp./Num. & 2D/3D & Bulk particles& $L/d$ & Walls. & Ref. \\ \hline
VC1& Exp. & 2D & aluminium cylinders, 60mm long, mixture of d=2 and 3mm & $ 16 \rightarrow 45$ & Plastic cylinders \hspace{5mm} $d_{\rm w}=2.5$ mm & \cite{PG96}\\ \hline
VC2 & Num (CD) & 2D & Disks, $e=0$, $\mu_p=0.4$ & 45 & Disks \hspace{9mm} $d_{\rm w}= 0.5d \rightarrow 4d$ &\cite{P02} \\ \hline
VC3 & Exp (MRI) & 3D & mustard seeds $d=1.3 {\rm mm}$& 42 & mustard seeds $d_{\rm w}=1.3{\rm mm}$&\cite{CPMDBGCR01} \\ \hline
\end{tabular}
\ec
\caption{Data sources for vertical chute flow. CD is for contact
dynamics, MRI for magnetic resonance imaging.}
\label{tablechute}
\end{table*}

\subsection{Kinematic properties}

\subsubsection{Velocity profile}

Typical velocity profiles obtained in quasi-static regime are plotted in figure~\ref{FigChuteFlow}b. The velocity is rescaled by the maximum velocity in the centre of the channel. In both experiments and numerical simulations, the profiles are characterised by a plug region in the centre part of the channel where the velocity is constant and the material not sheared. Variation of the velocity is localised in two shear zones close to the rough walls. The thickness of the shear
zones is of the order of 5 to 10 particles diameters in 2D or 3D, both in experiments and simulations. 

In some specific cases, intermittent flow occurs \cite{BGRH03} or, for much larger flow rates, density waves are observed \cite{BGRHHH02}, which might be related to the role played by the air trapped between the particles.

\subsubsection{Volume fraction profile}

Figure~\ref{FigChuteFlow}c shows typical volume fraction profile $\Phi(y)$ measured in simulations and experiments. In both cases the material is slightly less compact in the shear zone. In numerical simulations carried out with slightly polydispersed material, layering is observed as a consequence of the order induced by the walls.

\subsubsection{Velocity fluctuations profile}

In figure~\ref{FigChuteFlow}d the velocity fluctuations profile measured in 2D numerical simulations is plotted. It is rescaled by its mean value across the channel. We observe that the velocity fluctuates more in the shear zones close to the walls than in the plug region.

\subsection{Parametric study}

The existence of shear zones close to the wall in the vertical chute flow configuration is a very robust observation. It is then interesting to study the influence of the parameters of the problem on the thickness of the shear zones.

In figure~\ref{FigChuteFlow}e, we have plotted the experimental measurement obtained in 3D experiments for different flow rates. As expected in a quasi-static regime, the rescaled velocity profile and subsequently the width of the localised shear are independent of the flow rate.

Figure~\ref{FigChuteFlow}f displays velocity profiles obtained for different channel width $L$. The interesting result, also observed in other experiments~\cite{NL80}, is that the thickness of the shear zone does not vary much with $L$. It means that the relevant lengthscale that determines the shear zone is the particle diameter and not the channel width as it would be in a Poiseuille flow.

Another parameter that can be changed is the roughness of the wall. In figure~\ref{FigChuteFlow}g we have plotted the velocity profiles obtained when changing the size of the particles glued at the walls. Clearly, increasing the roughness increases the shear zone.

Finally the last parameter one can change is the inclination of the chute. Interestingly, as shown in figure~\ref{FigChuteFlow}f changing the angle $\theta$ from vertical changes the thickness of the shear zones: the bottom one increases and the top one shrinks. This is attributed to the change in stress distribution that occurs when inclining the silo~\cite{PG96}.

One can conclude from the parametric study that the flow in vertical channel develops localised shear zones close to the wall, whose thickness scales with the particle diameter. Changing the roughness or inclining the silo are the two main ways to change the shear zone thickness.

\section{Flow on inclined plane}
\begin{figure*}[ht!] \vspace{5 mm}
\bc
\includegraphics{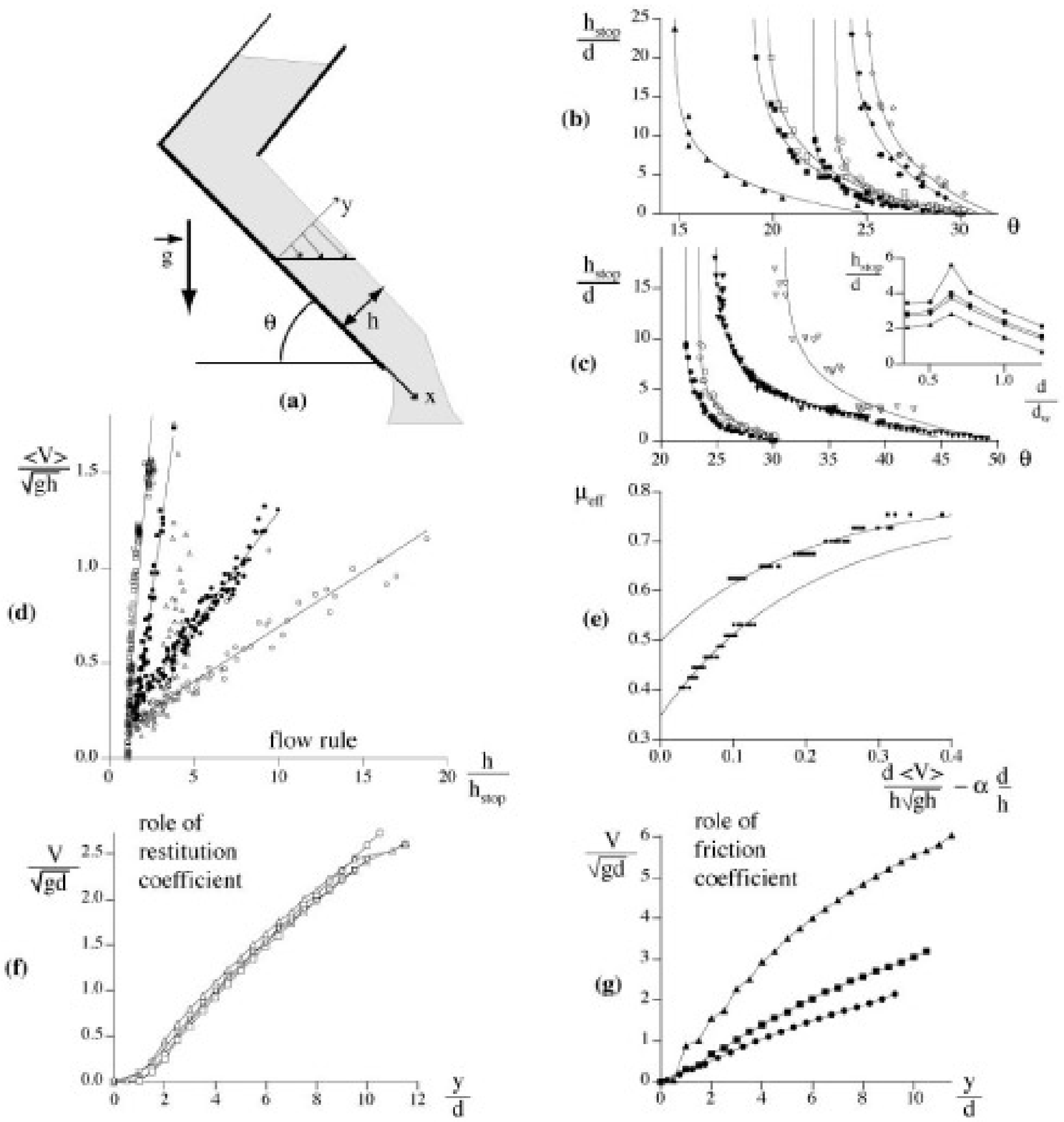}
\caption{
{\bf Rough inclined plane.}
{\bf (a)} Set up.
{\bf (b)} $h_{\rm stop}(\theta)$ (black symbols) and $h_{\rm start}(\theta)$ (white
symbols) from simulations IP3 ($\blacktriangle$), experiments IP5 with
glass beads ({\Large $\circ$},{\Large $\bullet$}), IP6 with mustard seeds ($\lozenge$, $\blacklozenge$) or with glass beads on carpet ($\square$, $\blacksquare$). See text for the fit.
{\bf (c)} Same as (b) for experiments IP5 with glass beads ({\Large $\circ$},{\Large $\bullet$}), IP7
with glass beads on velvet ($\triangledown$, $\blacktriangledown$). Inset: $h_{\rm stop}$ for different roughness condition from experiment IP8
$\theta=27^\circ$ ({\Large $\bullet$}), $\theta=28^\circ$ ($\blacksquare$),
$\theta=28.3^\circ$ ($\blacktriangledown$), $\theta=30^\circ$ ($\blacktriangle$).
{\bf (d)} Froude number $\langle V \rangle /\sqrt{gh}$ as a function of
$h/h_{\rm stop}(\theta)$ from simulation IP3 ({\Large $\circ$}
), experiments
IP5 with glass beads on glass beads ({\Large $\bullet$}), sand on sand ($\square$), IP6
sand on moquet ($\blacksquare$), IP7 glass beads on velvet ($\triangle$). Lines are fits by eq.~(\ref{frhstop}).
{\bf(e)} Effective friction deduced from the flow rule (see text). Experiment IP5 with glass beads ({\Large $\bullet$}), with sand ($\blacksquare$). Continous lines are deduced form eq.~(\ref{mueffinclinedplane}) and fit of $h_{\rm stop}(\theta)$. 
{\bf (f)} Velocity profiles from simulations IP4 ($\theta=14.4^\circ$,
$\mu=0.$) for different restitution coefficients $e=0.4$ ($\triangle$), $e=0.6$ ($\triangledown$), $e=0.7$ ($\square$)
and $e=0.8$ ({\Large $\circ$}).
{\bf (g)} Velocity profiles from simulations IP4 ($\theta=18^\circ$, $e=0.6$) for different friction coefficients $\mu=0.$ ($\blacktriangle$),
$\mu=0.25$ ($\blacksquare$) and $\mu=0.5$ ({\Large $\bullet$}).
}
\label{FigInclinedPlane}
\ec
\end{figure*}
\begin{figure*}[ht!] \bc \vspace{35 mm}
\addtocounter{figure}{-1}
\includegraphics{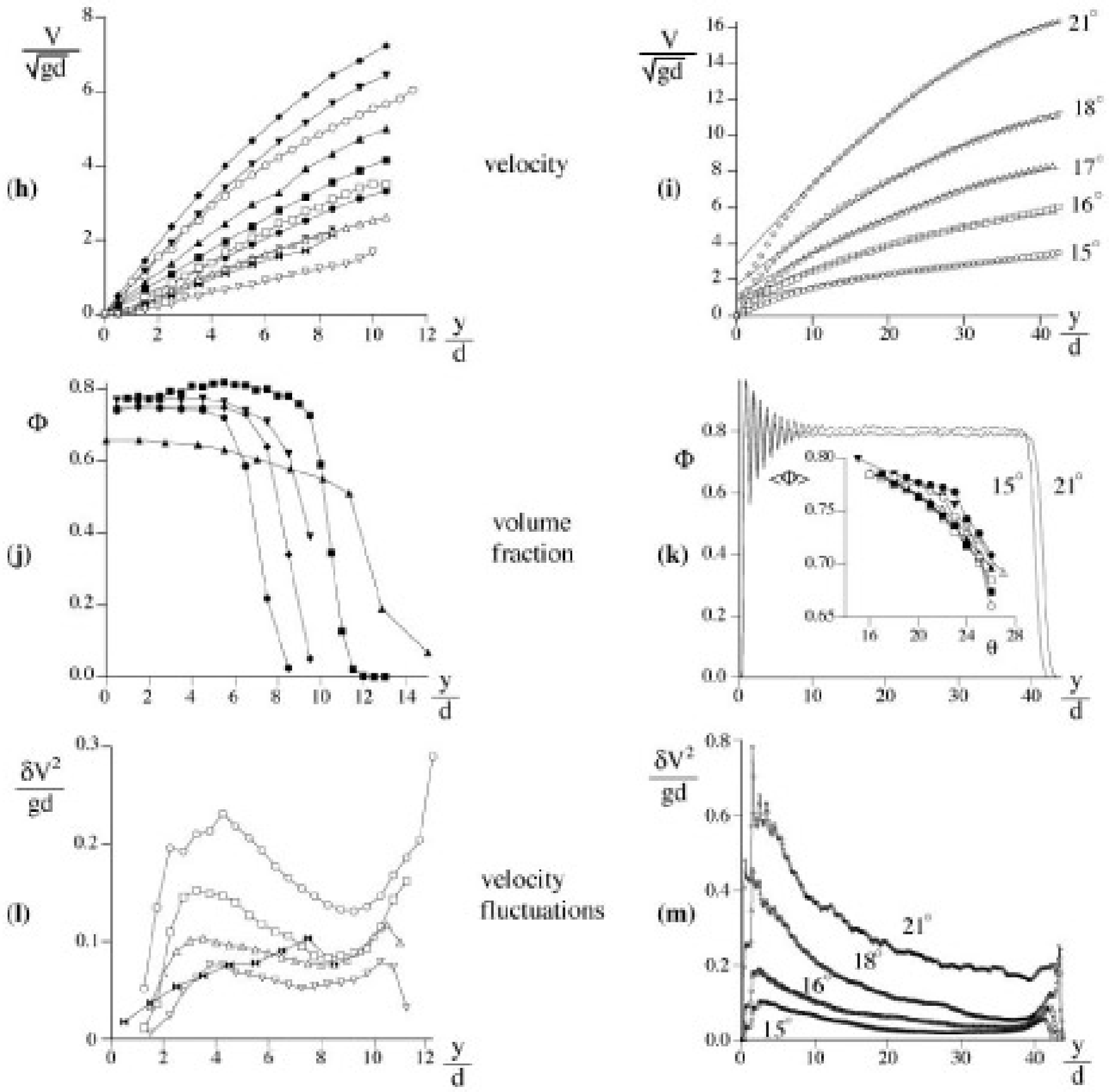}
\caption{
{\bf Rough inclined plane.}
{\bf (h)} Velocity profiles for thin flows from experiment IP2 for $\theta=34^\circ, 36^\circ$ (black
bow ties), IP1 for $\theta=21^\circ$ ({\Large $\bullet$}) ,
$22^\circ$ ($\blacksquare$), $23^\circ$ ($\blacktriangle$), $25^\circ$ ($\blacktriangledown$, $27^\circ$ ($\blacklozenge$), and simulations IP4 (e=0.6)
with $\theta=12.6^\circ$ ($\triangledown)$, $14.4^\circ$ ($\triangle)$, $16.2^\circ$ ($\square$),
$18^\circ$ ({\Large $\circ$});
{\bf (i)} Velocity profiles for thick flows from numerical simulations IP3.
{\bf (j)} Volume fraction profiles for thin layers from simulations IP4 ($e=0.7$) for $\theta=14.4^\circ $, ($\blacksquare$), IP3 for $\theta=23^\circ$ ($\blacklozenge$), from experiments IP2 for
$\theta=34^\circ$ ({\Large $\bullet$}), IP1 for $\theta=22^\circ$ ($\blacktriangledown$), IP9 for $\theta=27^\circ$ ($\blacktriangle$);
{\bf (k)} Volume fraction profiles for thick flows from simulation IP3; Inset: mean volume fraction versus inclination angle for different flow thickness;
{\bf (l)} Velocity fluctuations profiles for thin layers from experiment IP2 for $\theta=34^\circ$ (black bow ties) and from simulations IP4 ($e=0.6$) for $\theta=12.6^\circ$ ($\triangledown$), $\theta=14.4^\circ$ ($\triangle$), $\theta=16.2^\circ$ ($\square$) and $\theta=18^\circ$ ({\Large $\circ$});
{\bf (m)} Velocity fluctuations profiles for thick layers from simulations IP3.
}
\ec
\end{figure*}

\subsection{Set-up.}
The flows of granular material down an inclined plane are encountered in both geophysical and industrial contexts. The configuration (see Fig.~\ref{FigInclinedPlane}a) consists in a rough bottom inclined at an angle $\theta$ from horizontal. In experiments, both 2D and 3D, the granular material flows out from a reservoir located at the top of the plane. The flow rate is controlled by the opening of a gate. A dense granular flow then develops from the outlet. We will not discuss the case of rapid and dilute flows obtained when injecting the granular material from a hopper located far above the plane~\cite{JNJ90,A01,A97}. The bottom of the inclined plane is made of different materials: glued grains~\cite{A97,BDRTV03,P02,D98,P99,GTD03,FT04}, carpet~\cite{Ro03} or velvet cloth~\cite{DD99b,QA03}. In numerical simulations, periodic boundary conditions are imposed along the flow direction. The rough bottom is made of fixed particles. The simulation reported here are 2D only, but recent heavy computations have allowed 3D geometry \cite{SEGHLP01,SLG03}. All these configurations are reported in table~\ref{tableincl}.

In a given range of parameters that will be discussed below, a steady uniform flow of thickness $h$ is obtained. In this case, assuming a constant density $\rho$, the force balance leads to the following stress distribution : $\sigma_{xy}=\rho g \sin \theta (h-y)$, $\sigma_{yy}(y)=\rho g \cos \theta (h-y)$ and $\sigma_{xx}(y)$ remains undetermined.

\begin{table*}[ht!]
\bc
\begin{tabular}{|c|c|c|p{5.1cm}|c|p{4cm}|c|}
\hline
$\#$ & Exp./Num. & 2D/3D &material and plane &h/d & Walls & Ref \\ 
& & & & & & \\ \hline
IP1 & Exp.& 2D &aluminium beads: $d=3\mbox{ mm}$ $e=0.5$; plane: 2m long & $\approx 10$ & side walls: glass; bottom: glued grains & \cite{A97} \\ \hline
IP2 & Exp. & 2D & polystyrene disks: $d=8$\mbox{ mm} $e=0.4$; plane: $2$ m long &$\approx 10$ & side walls: glass; bottom: glued grains & \cite{BDRTV03} \\ \hline
IP3 & Num. (CD)& 2D & Disks: $e=0$, $\mu_{\rm p}=0.5$ & $\approx 50$ & side walls: none; bottom: glued grains & \cite{P02} \\\hline
IP4 & Num. (MD) & 2D & Disks: $e=0.4 \rightarrow 0.8$, $\mu_{\rm p}=0 \rightarrow 0.5$ &$\approx 10$ & side walls: none; bottom: glued grains & \cite{D98}\\ \hline
IP5 & Exp. & 3D & glass beads $d=0.5$ mm and sand $d=0.8$mm; plane: 2m long, 0.7m wide. &$< 20$ & side walls: none; bottom: glued grains & \cite{P99} \\ \hline
IP6 & Exp. & 3D & glass beads $d=1.5$ mm, sand $d=1$ mm, mustard seeds $d=2$ mm; plane: $2$ m long, $0.7$ m wide&$\approx 10$ & side walls: none; bottom: carpet& \cite{Ro03}\\ \hline
IP7 & Exp.& 3D & glass beads $ d=0.24$ mm; plane: $1.35$ m long, $0.6$ m wide&$\approx 10$ &side walls: none; bottom: velvet cloth& \cite{DD99b} \cite{QA03} \\ \hline
IP8 & Exp.& 3D & glass beads $d= 0.14 \rightarrow 0.53 $ mm; plane: $1.3$ m long, $0.6$ m wide&$< 20$ &side walls: none; bottom: glued grains& \cite{GTD03,FT04}\\ \hline
IP9 & Exp.& 3D & glass beads $d=1$ mm; plane $2$ m long, $0.05$ m wide& $ <100$ &side walls: Plexiglas with antielectrostatic film; bottom: glued grains & \cite{A01}\\ \hline
\end{tabular}
\ec
\caption{Data sources for inclined plane flow. MD stands for molecular dynamics simulations and CD for contact dynamics simulations}
\label{tableincl}
\end{table*}

\subsection{Flow threshold: transition between static and flowing states.}

An initially static granular layer of uniform thickness $h$ starts flowing when the plane inclination reaches a critical angle $\theta_{\rm start}$. Once initiated, the flow is sustained until the inclination is decreased down to a second critical angle $\theta_{\rm stop}$. The existence of these two angles is the evidence of the hysteretic nature of granular flows. In the case of inclined plane, these critical angles depend on the layer thickness $h$. Reciprocally, these thresholds can be interpreted in terms of critical layer thickness $h_{\rm stop}(\theta)$ and $h_{\rm start}(\theta)$. Indeed, the measurement of $h_{\rm stop}$ is easier as it corresponds to the thickness of the deposit remaining on the plane once the flow stops. The two curves $h_{\rm stop}(\theta)$ and $h_{\rm start}(\theta)$ divide the phase diagram $(h,\theta)$ in three regions: a region where no flow is possible ($h<h_{\rm stop}$), a subcritical region where both static and flowing layer can exist ($h_{\rm stop}<h<h_{\rm start}$) and a region where flow always occurs $h>h_{\rm start}$.

Measurements of these critical curves carried out for different materials and different rough bottoms are plotted in figures~\ref{FigInclinedPlane}b and~\ref{FigInclinedPlane}c. The curves $h_{{\rm stop}}(\theta)$ and $h_{\rm start}(\theta)$ exhibit the same shape for all the materials and can be fitted by:
$$h_{{\rm stop,start}}(\theta) / d=B \frac{\tan \theta_2-\tan \theta}{\tan \theta -\tan \theta_1}$$
where the fit parameters $\theta_1$, $\theta_2$ and $B$ depend on both the bulk material and the roughness conditions. As underlined in figure~\ref{FigInclinedPlane}c, changing the bottom rough plane from glued particles to velvet clothes dramatically shifts the critical curves $h_{\rm start}$ and $h_{\rm stop}$ toward higher angles.

In the flowing regime i.e. when $h>h_{\rm stop}(\theta)$, the flow is steady and uniform for moderate inclination, but accelerates along the plane for too large inclinations \cite{P02,A97,SEGHLP01,SLG03}. In the following we concentrate on the steady and uniform regime.

\subsection{Kinematic properties.}

\subsubsection{Velocity profiles}

Figures~\ref{FigInclinedPlane}h and~\ref{FigInclinedPlane}i respectively display the velocity profiles for thin and thick flows for different inclinations. In both cases, increasing $\theta$ increases the average shear rate and leads to more and more concave profile. Closer inspection of these profiles~\cite{P02} and recent numerical analysis for 3D flows~\cite{SEGHLP01,SLG03} reveal that for flow parameters ($h$,$\theta$) far enough from the flowing threshold curve $h_{\rm stop}(\theta)$ the velocity roughly obeys a Bagnold like profile:
\begin{equation}
\frac{V(y)}{\sqrt{gd}}=A(\theta)\frac{\left(h^{3/2}-(h-y)^{3/2} \right)}{d^{3/2}}
\end{equation}
\noindent
We will see in section~\ref{localreho} how one can extract the prefactor $A(\theta)$ from the bulk measurements, where Bagnold like rheology is valid. The continuous lines plotted in figure~\ref{FigInclinedPlane} i display the velocity profile obeying this rheology. One clearly see that the Bagnold profile fits the numerical data in the core region but not at the base nor at the free surface, where data exhibit a non zero shear rate. These regions of discrepancies apparently enlarge when inclination decreases. Close to the flowing threshold, for thin layers or low inclinations, the velocity profile becomes more linear (Fig.~\ref{FigInclinedPlane}h) \cite{SLG03}. Also, it is worth noting that for experiments IP9 carried out in a narrow channel~\cite{A01}, the velocity profiles can differ significantly from the above description and become convex. This observation reveals the role of the additional friction induced by lateral walls.

\subsubsection{Volume fraction profile}

The volume fraction profile $\Phi(y)$ is plotted in figures~\ref{FigInclinedPlane}j and~\ref{FigInclinedPlane}k for thin and thick layers. All the reported measurements show the same tendency: $\Phi(y)$ remains almost constant across the layer, except close to the free surface. This constant value appears to be independent of the flow thickness but decreases with the inclination as shown in the inset of figure~\ref{FigInclinedPlane}k. This behaviour is common to both experiments and numerical simulations~\cite{CPJM01,P02,SEGHLP01,SLG03}.

\subsubsection{Velocity fluctuation profile}

The velocity fluctuations $\delta V^2(y)$ are shown in figure~\ref{FigInclinedPlane}l and~\ref{FigInclinedPlane}m. Overall, for both thin and thick flows, $\delta V^2(y)$ increases with the inclination angle. The profiles exhibit two maxima, one close to the bottom, the other close to the surface. Both are of the same order for thin flows, whereas the fluctuations at the bottom dominate for thick flows. However, these features observed in numerical simulations do not show up in the only experimental measurements carried out with disk in 2D configuration (Fig.~\ref{FigInclinedPlane}l).

\subsection{Effective friction}

The inclined plane configuration gives information about the effective friction coefficient $\mu_{\rm eff}$ between the flowing layer and the rough bottom. The stress distribution for steady uniform flows implies that $\mu_{\rm eff}$ defined as the ratio between tangential and normal stress is simply equal to $\tan\theta$. Choosing an inclination for the plane is then equivalent to imposing the effective friction. The flow then adjusts its velocity so that the friction is equal to $\tan\theta$. One can then deduce how the effective friction evolves with velocity and thickness by measuring the flow rule : how does the mean velocity $\langle V \rangle$ of the granular layer varies with its inclination $\theta$ and thickness $h$? Figure~\ref{FigInclinedPlane}d shows experimental and numerical measurements of the relation $\langle V\rangle (\theta, h)$. The Froude number $Fr=\langle V \rangle/\sqrt{gh}$ is plotted versus the ratio $h/h_{\rm stop}(\theta)$. Each set of data collapse on a single curve indicating that the influence of the inclination seems to be encoded in the function $h_{\rm stop}(\theta)$. This correlation between flow velocity and deposit thickness is observed for different materials and different bottom coverage both in experiments and 2D simulations. Except for the experiments carried out with glass beads on velvet cloth (IP6), one observes the following scaling relation:
\begin{equation}
\frac{\langle V \rangle}{\sqrt{gh}}=\alpha+\beta\frac{h}{h_{\rm stop}(\theta)}
\label{frhstop}
\end{equation}
However, the coefficients $\alpha$ and $\beta$ are system dependent. Notice that for experiments using glass beads, $\alpha$ is zero. The same is observed in recent 3D simulations using spheres \cite{SLG03}. It is worth noting that the above relation may not be accurate in the vicinity of $h=h_{\rm stop}$. 

As $\mu_{\rm eff}=\tan \theta$, the effective friction coefficient
is obtained by inverting relation~\ref{frhstop} in order to express $\theta$ as a function of $\langle V \rangle$ and $h$. It is straightforward to show that according to relation~\ref{frhstop}, $\mu_{\rm eff}$ should be a function of a single parameter \cite{FP03} :

\begin{equation}
\mu_{\rm eff} \left(\langle V \rangle, h\right) =\mu_{\rm eff}\left(\frac{\langle V \rangle d}{h\sqrt{gh}}-\alpha \frac{d}{h} \right). 
\label{mueffinclinedplane}
\end{equation}

Figure~\ref{FigInclinedPlane}e shows the effective friction coefficient obtained by this procedure for two different materials. The continuous lines are $\mu_{\rm eff}$ functions extrapolated from eq.~(\ref{frhstop}) using fits of $h_{\rm stop}(\theta)$. In both cases, the effective friction coefficient increases when increasing the shear rate. Once again it is important to note that this relations is not valid for thickness close to the critical thickness $h=h_{\rm stop}$.

\subsection{Parametric study}

\subsubsection{ Dependence on the bottom roughness}

We have seen in the previous results that the roughness condition of the bottom plane strongly influences the flow properties. A systematic study has been carried out by gluing beads of diameter $d_{\rm w}$ and by changing gradually the flowing beads diameter $d$~\cite{GTD03}. In inset of figure~\ref{FigInclinedPlane}c, the deposit thickness $h_{\rm stop}$ is plotted versus the beads diameter ratio $d/d_{\rm w}$ for different inclinations. This work points out the existence of a given ratio $d/d_{\rm w}$ for which the deposit is maximum, which might correspond to a maximum of effective bottom friction. This ratio, independent of $\theta$, is mainly determined by the surface fraction of glued beads on the bottom plane~\cite{GTD03}.

\subsubsection{Influence of particle interaction parameters}

Such studies are essentially carried out in numerical simulations where one can independently vary the internal coefficient of friction $\mu_{\rm p}$ or the restitution coefficient $e$. In figure~\ref{FigInclinedPlane}f, we have plotted velocity profiles for the same $\theta$ and $h$ but for different coefficient of restitution $e$. The interesting result is that in the range $e < 0.8$, the profiles do not depend on $e$. This is to be contrasted with what would be expected in a kinetic regime dominated by binary collisions. The dependence on the friction coefficient $\mu_{\rm p}$ is also weak as shown in figure~\ref{FigInclinedPlane}g. Decreasing $\mu_{\rm p}$ slightly increases the values of the velocities \cite{A97}. However, choosing $\mu_{\rm p}=0$ seems to increase more dramatically the velocities. 

\section{Surface flows: heap flow and rotating drum}
%
\begin{figure*}[ht!] \bc
\includegraphics{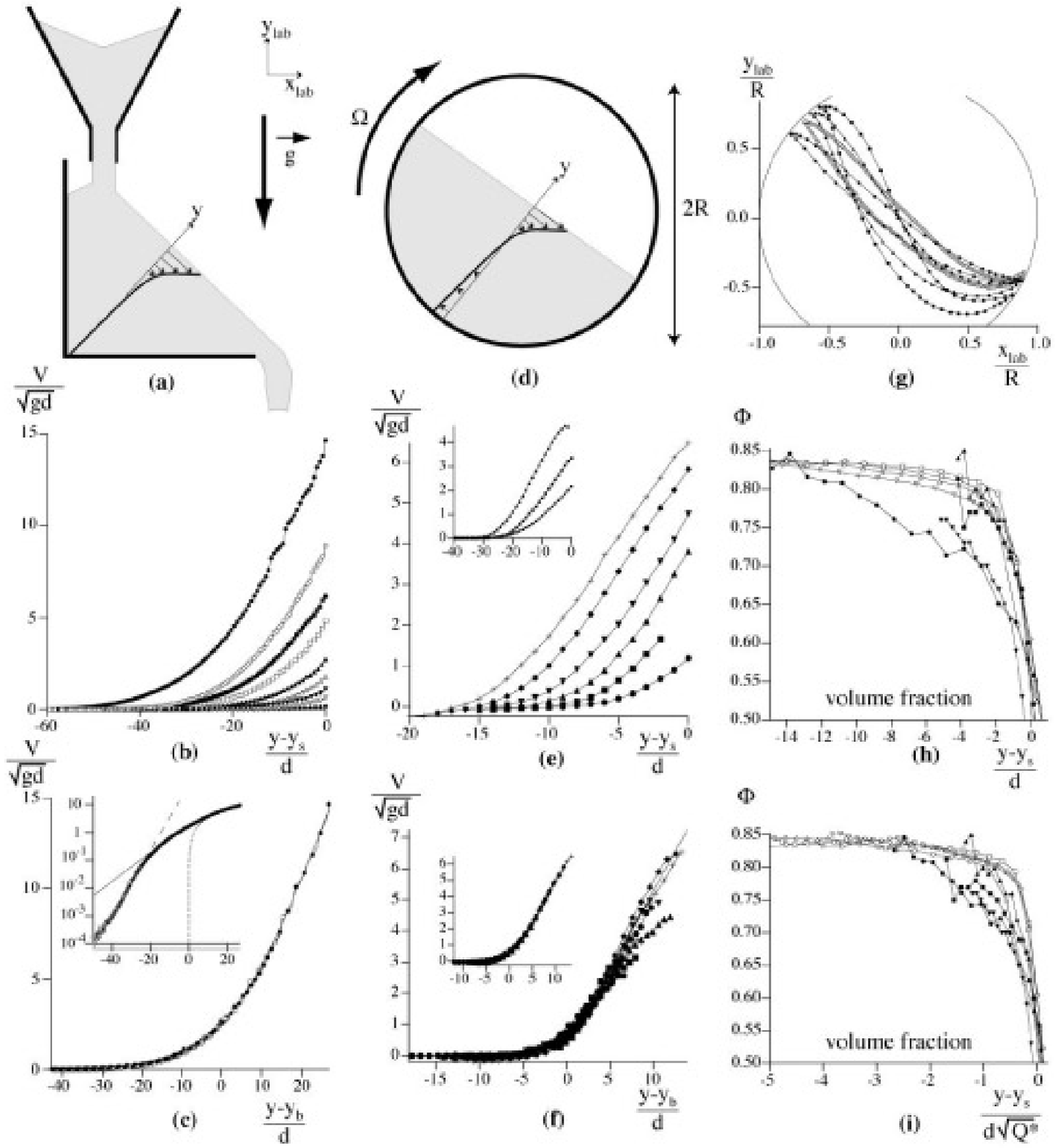}
\caption{
{\bf Heap and Drum surface flows.}
{\bf (a)} Heap set up.
{\bf (b)} Velocity profiles for different flow rates in the heap configuration from exp. SF4 with d=0.15 mm.
{\bf (c)} Same as (b) plotted as a function of $(y-y_b)/d$ (the y origin at the boundary between linear and exponential profile). Inset: Lin Log of the fastest profile. The fits are exponential (dashed line), linear (dotted line) and $\gamma h_e \ln (1+\exp (y/h_e))$ (solid line).
{\bf (d)} Rotating drum set up.
{\bf (e)} Velocity profiles from exp. 
SF6: $D/d=150$, $w/d=4.66$; from bottom to top $Q^*=5.25, 10.5, 15.75,
21, 30.75, 41.25$. Inset: velocity profiles from 2D CD simulation SF11, $D/d=150$
and $Q*=21, 30.75, 51.75$.
{\bf (f)} Velocity profiles as a function 
$(y-y_b)/d$ for different flow rates, different materials, in
different set up: steel beads with $D/d=150$, $w/d=7.33$
and $Q^*=5.25, 10.5, 15.75, 21, 30.75, 41.25$ from SF6; steel beads
with $D/d=133$, $w/d=1$ and $Q^*=6, 10, 18$ from SF5; glass beads with
$D/d=100$, $w/d=40$ and $Q^*=7.5, 17, 26, 36.5$ from SF7. Inset: same plot with data from SF6 only.}
\label{FigSurfaceFlow}
\ec
\end{figure*}

\begin{figure*}[ht!] \bc \vspace{25 mm}
\addtocounter{figure}{-1}
\includegraphics{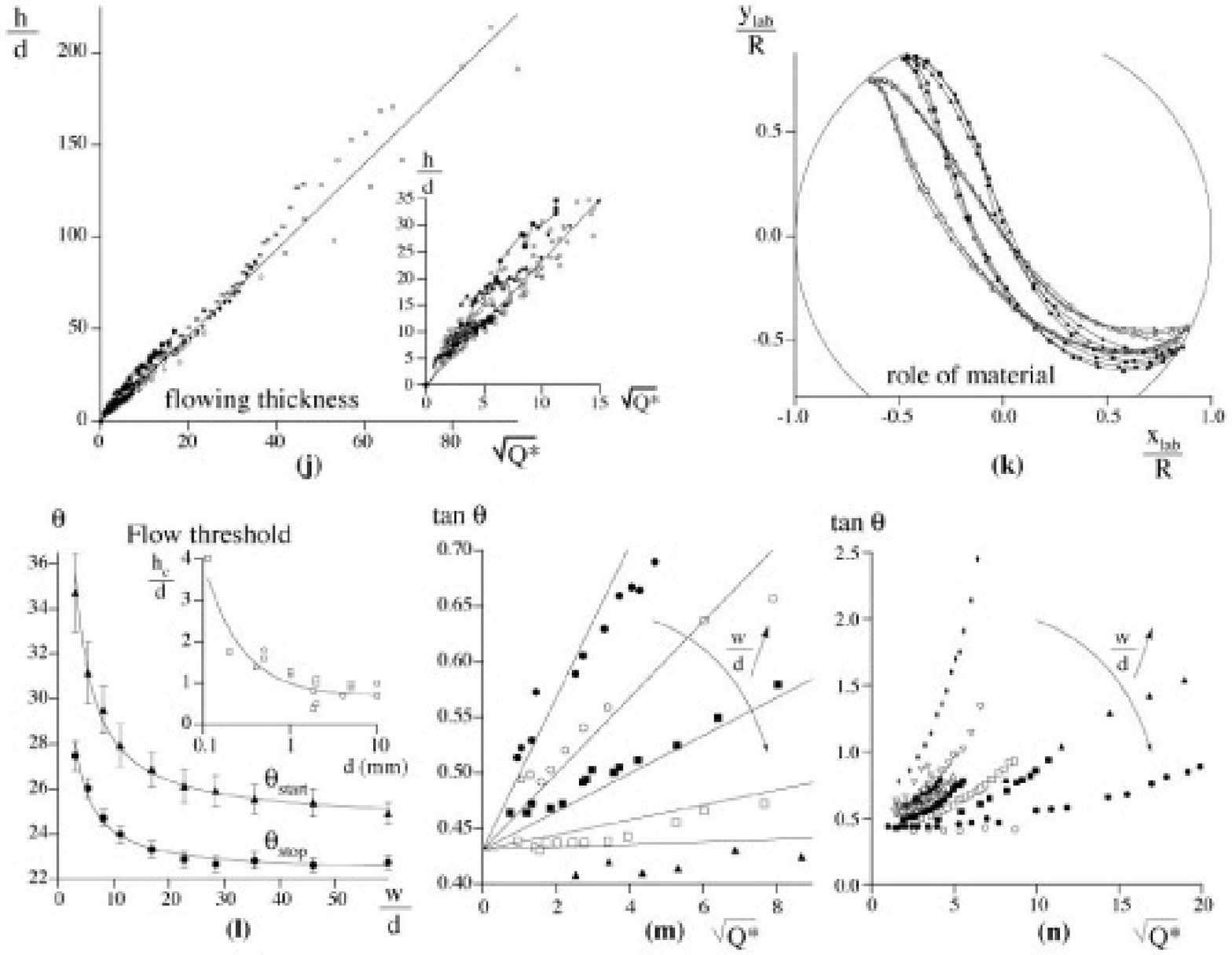}
\caption{
{\bf Heap and Drum surface flows.}
{\bf (g)} Shape of the flowing layer (free surface $y_s$ and
bed-layer interface $y_b$) observed with steel beads. Black labels
are from SF10: $gap/d=5$, $D/d=160$ with $Q^{*}=16.0; 52.8; 94.4$.
White labels are from SF6: $gap/d$=2.33, $D/d=150$ with $Q^{*}=30.75$.
{\bf (h)} Volume fraction profiles as a function of
$(y-y_s)/d$ obtained with steel beads with
 $D/d=133$, $w/d=1$ and $Q^*=6.0, 9.7, 18.1$ from SF5 (dark labels),
and with white labels, $D/d=150$, $w/d=1$ and $Q^*=31.0$ from SF6, 
numerically with steel beads, $D/d=150$ and $Q*=21, 30.75, 51.75$ from
SF11.
{\bf (i)} Same profiles as for (h) but plotted as a function of $(y-y_s)/d$ rescaled by $\sqrt{Q^*}$.
{\bf (j)} Flowing layer thickness $h/d$ at the centre of the cell as a function of $\sqrt{Q^*}$. Inset: is a zoom on small $Q^*$. Data from drum experiments SF2, SF5, SF6, SF7, SF8, SF9, SF10, with different materials -sand, glass, steel-- with different drum size, $D/d=[40-2500]$ and different gap size $w/d=[2.33-610]$. The points linked together are from heap experiments SF4. 
{\bf (k)} Rescaled free surface and bed-layer interface from SF8, SF9, SF10 at two different flow rate, with different materials (the layers are rotated relatively to each other to match their dynamical angle of repose). Upper curves: steel beads ($d=1mm$; $d=2mm$) and glass beads ($d=2mm$); $D/d=160$ and $Q^{*}=52.8$. Lower curves: sand ($d=0.4mm$ and $d=0.8mm$) and glass beads ($d=0.8mm$); $D/d=400$ and $Q^{*}=210$.
{\bf (l)} Critical angles $\theta_{\rm start}$ ($\blacktriangle$) and $\theta_{\rm stop}$ ({\Large $\bullet$}) as a function of the gap width $W$ for glass beads $d$=1.85 mm. Solid lines correspond to fit by eq.(\ref{EqParois}). Inset: Characteristic length scale
$h_{\rm c}$ of wall effect as a function of
$d$.
{\bf (m)} Free surface slope $\theta$ as a function of $\sqrt{Q^*}$ for the heap experiment; data from SF3 with $w/d \in [10-610]$; the lines corresponds to the approximation $\tan \theta=\mu_\infty+\mu_w \sqrt{Q^*} d/w$.
{\bf (n)} Free surface slope $\theta$ as a function of $\sqrt{Q^*}$ in rotating drums same data as in (j).}
\ec
\end{figure*}

Granular flows confined to a surface layer on a static granular bed are probably the most frequently encountered in industrial process and nature. Accordingly, they have been extensively studied in the past for practical interest and more recently as model system in fundamental studies~\cite{R00,DADC01,R90,LJN91,GH97,ZXY01,ZXYZ02,ER88,NACFJ93,CDFL95,KMSO97,EV98,DRMN98,YNATT98,KINN01,KOAO01,OK01,BDL01,BDL01b,F02,C03,R03}.

\subsection{Set-up}
Most experimental work has been conducted in two systems: down a heap~\cite{LJN91,GH97,DADC01,ZXY01,ZXYZ02} and inside the so-called rotating drum~\cite{R00,R90,ER88,NACFJ93,CDFL95,KMSO97,EV98,DRMN98,YNATT98,KINN01,KOAO01,OK01,BDL01,BDL01b,F02,C03,R03}, both shown schematically on figure~\ref{FigSurfaceFlow}a and \ref{FigSurfaceFlow}d. The flow down a heap is most commonly obtained in a Hele Shaw cell: beads are poured in between two glass plates separated by a distance $W$. The flow rate per unit of width $Q$ is controlled by the hopper outlet. After a transient stage, the cell is full and one obtains a stationary regime with equilibrated fluxes at the pouring point and at the exit of the cell. The rotating drum of width $W$ and diameter $2R$, is half-filled with the grains and rotated at constant angular velocity $\Omega$. For an appropriate range of angular velocity, one obtains a stationary flowing layer, with a given flow rate per unit of width $Q=\Omega D^2/8$, where $D$ is the drum diameter. In both cases, the rescaled flow rate
\begin{equation}
Q^*=\frac{Q}{d \sqrt{gd}}
\end{equation}
is the unique parameter controlling the flowing layer thickness $h$, the angle of the free surface $\theta$, once the geometrical parameters $D/d$ and $W/d$ are fixed.

The major advantage of the heap geometry is that it easily produces homogeneous flows. However, it is difficult to explore a wide range of flowing layer thickness and surface inclination. Conversely, the rotating drum set up allows to explore stationary flows in a much broader range of both $h$ and $\theta$ but the flow is not strictly homogeneous in the flowing direction. Still, for drums of large enough diameter, one expects the flow at the centre of the drum to be independent of the drum size, which we will discuss further in the light of the experimental data. Under this assumption of uniformity, the stress distribution in the central part of the drum is the same as in the inclined plane case, that is $\sigma_{xy}=\rho g \sin\theta(y_s-y)$ and $\sigma_{yy}=\rho g \cos \theta (y_s-y)$, where $y_s$ is the free surface coordinate. The table \ref{tabledrum} summarises the various flow data sources used in the present section.

\begin{table*}[ht!]
\bc
\begin{tabular}{|c|c|c|l|c|c|c|c|}
\hline
 $\#$ & Exp/Num & 2D/3D & particles material & Q* & W/d & D/d & Ref \\ \hline
 SF1 & Exp Heap & 2D\&3D & glass; $d=0.1$mm & $<1$ & $1\rightarrow 10$ & &\cite{GH97} \\ \hline
 SF2 & Exp Drum & 3D & glass; $d=0.23\rightarrow 3$ mm & $<1$ & $10\rightarrow 90$ &$60-700$ &\cite{C03,CGPR03b} \\ \hline
 SF3 & Exp Heap & 3D & glass $d=0.5$ mm & $1\rightarrow 75$ & $10\rightarrow 610$ & &\cite{P03} \\ \hline
 SF4 & Exp Heap & 3D &glass $d=0.25$ mm and $d=0.15$ mm& $1\rightarrow 15$ & $120$ & &\cite{A03} \\ \hline
 SF5 & Exp Drum & 2D & steel; $d=1.5$ mm & $5\rightarrow 20$ & $1$ & 133 &\cite{R00}\\ \hline
 SF6 & Exp Drum & 2D\&3D & steel, aluminium; $d=3$ mm & $5\rightarrow 50$ & $1\rightarrow 7.33$ & 150 &\cite{B01,BDL01} \\ \hline
 SF7 & Exp Drum & 3D & glass; $d=0.2\rightarrow 2$ mm & $5\rightarrow 4500$& $2.5\rightarrow 120$& $50\rightarrow 2500$ &\cite{F02} \\ \hline
 SF8 & Exp Drum & 3D & glass; $d=0.8\rightarrow 4$ mm & $2\rightarrow 360$& $5\rightarrow 12.5$ & $40\rightarrow 400$ &\cite{OK01} \\ \hline
 SF9 & Exp Drum & 3D & sand; $d=0.4\rightarrow 0.8$ mm & $5\rightarrow 360$& $12.5\rightarrow 25$ & $5\rightarrow 400$ &\cite{OK01} \\ \hline
 SF10 & Exp Drum & 3D & steel; $d=1\rightarrow 4$ mm & $2\rightarrow 260$& $5\rightarrow 10$ & $40\rightarrow 320$ &\cite{OK01} \\ \hline
 SF11 & Num Drum & 2D & steel; $d=0.23\rightarrow 8$ mm & $5\rightarrow 50$ & & 150 &\cite{R03} \\ \hline
\end{tabular}
\caption{Data sources for Surface Flows (heap and rotating drum. Experiment SF1 and SF2 are dedicated to the study of avalanches, whereas experiment SF3 to SF10 deal with stationary flows. Nota Bene: simulation SF11 is real 2D, that is the particles used are infinite cylinders.}
\label{tabledrum}
\ec
\end{table*}

\subsection{Transition between static and flowing regimes}

For small flow rate, typically $Q^*<1$, intermittent avalanches occur~\cite{R00,DD99b,R90,ER88,CDFL95,C03,CGPR03}. As a result, the surface slope angle oscillates between the angle at which an avalanche is triggered $\theta_{\rm start}$ and the static angle of repose $\theta_{\rm stop}$ that remains after the avalanche. The presence of confining lateral walls is known to improve the stability of a pile~\cite{LJN91,GH97,ZXY01,B99}. For narrow channel (small $W$) the angles are higher. A typical evolution of both characteristic angles $\theta_{\rm start}$ and $\theta_{\rm stop}$ with $W$, obtained for glass beads in a rotating drum set-up with glass walls is displayed in figure~\ref{FigSurfaceFlow}l: both $\theta_{\rm start}$ and $\theta_{\rm stop}$ decrease with increasing $W$ towards asymptotic values for large gap widths. These evolutions are well described by the equation:
\begin{equation}
\label{EqParois}
\tan \theta_{\rm{start,stop}} \,=\, \tan
\theta_{\rm{start,stop}}^{\infty} \, + \frac{h_{\rm c}}{W},
\end{equation}
where $\theta_{\rm{start,stop}}^{\infty}$ are the asymptotic values of $\theta_{\rm{start,stop}}$ and $h_{\rm c}$ the corresponding characteristic length scale of wall effect. This equation is physically consistent with additional friction forces induced by the walls~\cite{C03,CGPR03b}. The inset of figure~\ref{FigSurfaceFlow}l displays the characteristic length $h_{\rm c}$ as a function of the bead diameter $d$ obtained with equation (\ref{EqParois}) for all data sets found in the literature involving glass beads in various setups. It clearly puts in light two different regimes depending on the bead diameter $d$. Whereas $h_{\rm c}$ is proportional to $d$ for large beads ($d>0.5{\rm mm}$), which implies that wall effect is a geometric effect, $h_{\rm c}$ is constant for small bead diameters ($d<0.5{\rm mm}$). One explanation to this constant value, independent of $d$, could be that small beads aggregate because of surface forces such as van der Waals forces~\cite{C03,CGPR03b}.

\subsection{Kinematic properties}

Figure~\ref{FigSurfaceFlow}g displays the typical shape of the flowing layer in the rotating drum, when the flow is stationary -- in practice when $Q^*>1$ -- with the characteristic S-shape of the free surface $y_s(x)$ and the essentially convex shape of the bed-layer interface $y_b(x)$. The maximum layer thickness $h=y_s-y_b$ increases and the free surface becomes more S-shaped with increasing $Q^*$. Accordingly the slope in the centre of the drum is also accentuated. Qualitatively, similar observations are made with all materials. Obviously the flow is not homogeneous, apart from the centre of the drum where both the layer thickness and the local slope of the interfaces vary slowly along the interface, for large enough drum $D/d>50$. Accordingly, in the drums, kinematic profiles have always been measured in the centre of the drum. In the heap, profiles has been measured at in the centre of the cell in between the hopper and the outlet. 

\subsubsection{Velocity profile}

Figure~\ref{FigSurfaceFlow}b and \ref{FigSurfaceFlow}e displays typical velocity profiles obtained respectively in heap \cite{A03} and drum configuration~\cite{BDL01}. The similarity between both configurations is striking. In both cases, the profiles are localised under the free surface and are composed of an upper linear part in the flowing layer and a lower exponential tail in the granular bed. The exponential tail is clearly evidenced in inset of figure~\ref{FigSurfaceFlow}c. The crossover between these two behaviours extends over a wide zone of approximately ten grains. An interface between the flowing layer and the quasi static pile can be defined extrapolating the linear part of the velocity profile to zero. One can plot the profiles putting the $y$ origin at this interface as done in figures \ref{FigSurfaceFlow}c and \ref{FigSurfaceFlow}f. In both geometries, one observes a collapse indicating that a universal profile exists. By increasing the flow rate, one simply explores a wider and wider zone of this profile. The shear rate $\dot \gamma$ in the linear part --when it exists-- is essentially constant, independent of the flow rate and in both geometries equal to :

\begin{equation}
\dot\gamma \simeq 0.5 \sqrt{\frac{g}{d}}.
\label{shearrate}
\end{equation}
Still, in recent real 2D numerical results in the rotating drum geometry~\cite{R03}, under the same condition as in~\cite{BDL01}, the flowing layer appears twice deeper and the linear profiles exhibit a dependence of the shear rate with $Q^*$ (inset of figure~\ref{FigSurfaceFlow}e). This preliminary result which has to be confirmed, might indicate some non trivial effect of the wall friction.

The thickness of the flowing layer $h=y_s-y_b$ has been measured in all the experiments. Fig~\ref{FigSurfaceFlow}j demonstrates the existence of a general scaling
\begin{equation}
\frac{h}{d} \propto \sqrt{Q*}
\end{equation}
for all the data collected in surface flows, heap and drum, with different materials -- glass, sand, steel -- and very different drum size ranging from $40$ to $2500$. As a matter of fact, assuming a purely linear velocity profile in the flowing layer, the above scaling is exactly equivalent to the existence of a constant shear rate of the order of $0.4 \sqrt{\frac{g}{d}}$. This is slightly smaller than the shear rate measured directly from the velocity profiles (eq.~\ref{shearrate}).

\subsubsection{Compacity profile}

As shown on figure~\ref{FigSurfaceFlow}h the compacity -- here measured in 2D configurations -- increases across the layer from $0.6$, at the free surface to $0.8$, its close random packing value, at the bed-layer interface. This behaviour is common to both experiments and numerical simulations. The compacity seems to decrease on a typical scale $\sqrt{Q^*}d$ as shown in figure~\ref{FigSurfaceFlow}i.

\subsection{Effective friction}

As for the flow down an inclined plane, the stress distribution in the flowing layer is such that the effective friction coefficient is again the tangent of the dynamical angle of repose.

In the case of the flow down a heap (see Fig.~\ref{FigSurfaceFlow}m), the pile slope increases linearly with $\sqrt{Q*}$. This dependency becomes weaker when the channel width increases and for the widest channel the pile slope even becomes independent of the flow rate. This suggests that the increase of the effective friction with flow rate is purely induced by the additional wall friction. Accordingly, one can propose a single fit to describe this wall effect:

\begin{equation}
\tan(\theta)=\tan(\theta)_{\infty}+\mu_{\rm w} \sqrt{Q^*}\frac{d}{W} , 
\end{equation}
where $\sqrt{Q*}$ is again interpreted as the dimensionless thickness of the flowing layer experiencing the wall friction. This is reminiscent from eq.~(\ref{EqParois}) where the length $h_{\rm c}$ would be equal to $\mu_{\rm w} \sqrt{Q^*_c} d$. The critical flow rate $\sqrt{Q^*_c}$ is then equal to 1.5 and can be interpreted as the flow rate below which no permanent flow can be sustained. 

In the case of the rotating drum, a similar general tendency is observed on figure~\ref{FigSurfaceFlow}n. For a given drum width, the pile slope increases with the flow rate although facing the different set of data, one can hardly conclude to the linear dependency with $\sqrt{Q^*}$. Also, the pile slope becomes less sensitive to the flow rate for wider drums despite some discrepancies across the numerous experiments.

Altogether, these results lead us to conjecture that in the limit of infinite gap, the effective friction is constant, independent of $Q*$ and that the observed dependencies are related to intricate wall and geometrical effects.

\subsection{Parametric study}

Except for the pile slope, the materials intrinsic properties seem to have very little effect on the flow properties in a rotating drum. The velocity gradient inside the flowing layer is identical for three different materials (glass, steel and aluminium) as observed on figure~\ref{FigSurfaceFlow}e. Also, the relation between the flowing thickness and the flow rate is the same for different materials (sand, glass and steel) as shown on figure~\ref{FigSurfaceFlow}j.

Finally, it is worth noting that the independence of the flowing layer thickness with the material extends to non-uniform flows. This is very well demonstrated on figure~\ref{FigSurfaceFlow}k, where the free surface and the bed-layer interface for different materials and different drum size are plotted in a frame scaled by the drum radius, and rotated so that the dynamical angles of repose coincide.

\section{Discussion}

In the above sections we have gathered data for each of the flow configurations, obtained with different experimental or numerical conditions. Doing so, we were able to identify the relevant flow characteristics in the different geometries. In order to identify simple and basic features underlying common physical phenomena, we now review the common features and differences arising among the configurations.

A first general observation is that the driving force must overcome some "static" threshold -- the yield stress -- in order to enable a dense granular flow. Once the flow is running, it can be sustained for driving forces lower than this "static" threshold resulting into a hysteretic behaviour. This has been clearly evidenced in the annular shear cell (Fig.~\ref{FigAnnularShear}b), on the inclined plane (Figs.~\ref{FigInclinedPlane}b and \ref{FigInclinedPlane}c) as well as in the heap -- or drum -- geometry (Fig.~\ref{FigSurfaceFlow}l). In the other geometries, the data we have correspond to flows where the deformation is imposed, whereas hysteresis is observed when the stresses are imposed. The static threshold usually depends on the history of the sample \cite{VHCBC99}. Also, the static and dynamical friction coefficients depend on the material mechanical properties. But the latter do not strongly influence the kinematic properties of the flows.

In all the geometries, a dense flow regime - either quasi-static or inertial - was identified, separated from the dilute collisional regime. A striking feature is the diversity of velocity profiles observed in the different geometries. When the flow is confined (annular shear, chute flow) the shear is localised close to the driving wall and the velocity decreases over few grain sizes. However, in the perfect plane shear, no localisation is observed and the velocity profile remains linear. In the case of free surface flows, the velocity follows either Bagnold or linear profiles in the case of the inclined plane geometry (Figs.~\ref{FigInclinedPlane}h and \ref{FigInclinedPlane}i), whereas it is always linear with an exponential tail in the drum and heap cases (Fig.~\ref{FigSurfaceFlow}b and~\ref{FigSurfaceFlow}e). There are two interesting connections between these three types of profile (exponential, linear and Bagnold). First, flow on a heap present simultaneously (one on the top of the other) the dense inertial flow (linear profile) and the quasi-static confined flow (exponential profile) (Fig.~\ref{FigSurfaceFlow}c). Second, the flow on an inclined plane exhibits a Bagnold profile (Fig.~\ref{FigInclinedPlane}i) in the limit when the flowing height $h$ is large compared to the critical layer thickness $h_{\rm stop}$. As soon as $h$ becomes comparable to $h_{\rm stop}$, the profile becomes linear (Fig.~\ref{FigInclinedPlane}h). This suggests that a continuous transition between the inclined plane flow and the surface flow could exist. Finally, it is worth underlining that the velocity fluctuations, when they were measured, seem to be strongly related to both the local shear rate and pressure (Figs.~\ref{FigPlaneShear}g and~\ref{FigAnnularShear}h).

Although rudimentary, this transverse reading raises many questions. What are the relevant time and length scales in the different configurations? How does the transitions arise between the different flow regimes in the different configurations? Is a single rheometer - for instance the plane shear - sufficient to predict the kinematic properties in all the geometries? By comparing the data can we get information about the granular rheology? In the following we discuss in more details the similarities and differences arising from the different configurations. We first analyse the relevant length and time scales, before studying the different flow regimes and discussing some minimal rheological descriptions based on dimensional analysis. 

\subsection{Relevant parameters}

We wish to discuss first the relevant dynamical mechanisms and the corresponding parameters. For this, it is useful to distinguish three scales of very different natures: the microscopic scale at which the contact between grains is established, the grain level at which the different forces act and the scale of the flow itself (of the geometry) which determines the nature of the granular flow.

\subsubsection{Microscopic mechanisms at the contact scale}

The roughness of the grains at the microscopic scale is responsible for the contact friction between grains. The results presented here, for instance the numerical simulations of shear flows (Fig.~\ref{FigPlaneShear}), show no influence of this roughness on the kinematic properties of the flow. The same thing is demonstrated for the shape of the grains, by the rotating drum experiment (sand vs glass beads, Fig.~\ref{FigSurfaceFlow}k). These microscopic lengthscales only modify the effective friction coefficients (Figs.~\ref{FigPlaneShear}e, \ref{FigInclinedPlane}b, \ref{FigInclinedPlane}c, \ref{FigInclinedPlane}e and \ref{FigSurfaceFlow}k) and seem entirely encoded in them. It is striking to observe that the effective friction $\mu_{\rm eff}$ increases dramatically with the interparticle friction coefficient $\mu_{\rm p}$, near $\mu_{\rm p}=0$. We see on Fig.~\ref{FigPlaneShear}e that the effective friction coefficients measured for $0.1<\mu_{\rm p}<0.8$ almost collapse on the same curve but are far above those obtained for vanishing $\mu_{\rm p}$. The same tendency can be inferred from Fig.~\ref{FigInclinedPlane}g for the inclined plane experiment : the two velocity profile for non zero $\mu_{\rm p}$ are very close whereas the case $\mu_{\rm p}=0$ yields larger velocities. It would be interesting to study the behaviour of the effective friction $\mu_{\rm eff}$ as a function of $\mu_{\rm p}$ for very small $\mu_{\rm p}$ (with log-scale variations).

During a collision between grains, there is at the same time a contact force that push the grains back and a dissipation of energy related to the inelasticity. One can associate two different timescales to these two effects: the collision time, which is determined by the elastic properties of the grain and the dissipation timescale, given by the typical time for the internal elastic vibrations to be damped. In the simplest case of a collision between two grains, the effective restitution coefficient $e$ decreases with the ratio of the dissipation timescale to the elastic timescale. It is much more difficult to estimate these timescales as soon as one considers an assembly of grains in permanent contact. For instance, in a system of size $W$, the time needed for the elastic wave to propagate across the cell is in fact $W/d$ times the collision time \cite{D04}. The influence of these two parameters have not been investigated separately but only through $e$. As seen in the plane shear flow simulations, this restitution coefficient influences the transition from dense to dilute collisional regimes: more elastic particles have a "gaseous" behaviour for smaller shear rate \cite{IK02}. However, it turns out to have no influence on the granular flow itself (Figs.~\ref{FigPlaneShear}e, \ref{FigInclinedPlane}f and \ref{FigSurfaceFlow}k).

We can draw a general and solid conclusion: as soon as there is a separation between the flow timescales (see below) and the microscopic ones, the later have no influence on the flow characteristics. In other words, the system is equivalent to rigid inelastic spheres when both the energy dissipation and the elastic vibrations are much more rapid than the flow timescales. It is important to note that this limit becomes difficult to achieve in practice in the elastic limit $e \to 1$ since the dissipation timescale becomes much larger than the collision time. 

\subsubsection{Mechanisms at the grain level}
\begin{figure}[h!] \bc
\includegraphics{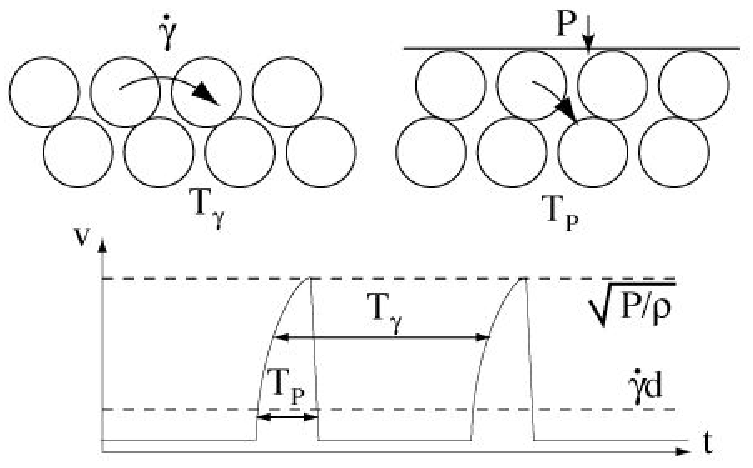}
\caption{Schematic showing the physical meaning of the typical time of deformation $T_{\gamma}$ and the confinement timescale $ T_{\rm p}$.}
\label{TimeScale}
\ec
\end{figure}

So, there is no influence of microscopic timescales on macroscopic flow properties. Accordingly, the grain size $d$ is the natural lengthscale of granular problems - except specific lengths related to the geometry (see below). As there is only one mass in the problem (that of the grain), granular flows are independent of the material density. In the homogeneous simple shear flows considered here, the strain tensor depends only on one parameter, the shear rate $\dot \gamma$ and the stress tensor on two parameters, the normal stress $P$ and the shear stress $\tau$. These three quantities define two independent dimensionless numbers, the effective friction coefficient,
\begin{equation}
\mu_{\rm eff}=\frac{\tau}{P}.
\end{equation}
and the rescaled shear rate
\begin{equation}
I=\frac{\dot \gamma d}{\sqrt{P/\rho}}
\label{defI}
\end{equation}
The parameter $I$ can be interpreted as the ratio of two different timescales at the grain level:
\begin{equation}
I=\frac{T_{\rm p}}{T_{\gamma}}.
\end{equation}
$T_{\gamma}$ is the typical time of deformation :
\begin{equation}
\label{Tgamma}
T_{\gamma}=\frac{1}{\dot \gamma}
\end{equation}
and $ T_{\rm p}$ is the confinement timescale:
\begin{equation}
\label{TP}
T_{\rm p}=d \sqrt{\frac{\rho }{P}}.
\end{equation}
Imagine two layers of grains moving one on top of the other, as shown in figure~\ref{TimeScale}. $T_{\gamma}$ is the macroscopic time needed for one layer to travel over a distance $d$ with respect to the other. $T_{\rm p}$ can be interpreted as the time needed by the top layer to be pushed back to its lower position, once it has climbed over the next particle. These two timescales can be very different, for instance in the case of quasi-static deformation. The motion is then made of a succession of a very slow motions when the particle climbs over the next one, and a rapid motion when it is pushed back into the next hole by the confining pressure. The typical velocity time evolution would be as drawn in figure~\ref{TimeScale}.
Within this simplistic picture, the volume fraction $\Phi$ is governed by the fraction of time during which the grains actually move. It thus suggests that $\Phi$ should be a slaved variable of $I$. In the following, we will assume that it is indeed the case. This is of course a strong hypothesis we shall reconsider in the following. So, if there exists a local unique rheology, there should be a unique relationship between the rescaled shear stress $\mu_{\rm eff}$ and the rescaled shear rate $I$.

$I$ can obviously be defined locally in all the situations but we can also estimate its typical value in the different geometries. In the annular shear, the pressure increases linearly with depth. When measurements are performed at the free surface, $P$ is of the order of $\rho gd$ so that $I$ can be defined as $\sqrt{d/g} \dot \gamma$ which is nothing but the rescaled shear rate (see Fig.~\ref{FigAnnularShear}h). When measurements are performed at half width, $P$ is of the order of $\rho gW/2$ so that $I$ can be defined as $\sqrt{W/2g} \dot \gamma$. In the chute flow, the pressure is limited to $P=\rho g L$ according to Janssen effect and $I$ can be defined locally as $\dot \gamma d/\sqrt{gL}$. In the free surface flows, the pressure increases with depth as $P=\rho g (h-y) \cos{\theta}$ so that its average value is $<P>=\rho g h \cos{\theta}/2$. The mean velocity gradient scales on $<V>/h$ so that the average of $I$ can be estimated by:
\begin{equation}
<I>=\frac{5}{2} \; \frac{<V>d}{h\sqrt{gh\cos{\theta}}}
\end{equation}
where the factor $5/2$ is derived assuming a Bagnold like velocity profile (see below).

\subsubsection{Geometrical parameters}
If the dimensionless number $I$ characterising the relative importance of inertial with respect to confining effects can be defined in all the cases, there are other parameters that are specific to a geometry i.e. that characterise the boundary conditions. One can see for instance that the presence of side walls (surface flow) or a bottom rough boundary (inclined plane) or both (chute flow) have a strong impact on the flow itself. Most of the geometrical aspect ratios turn out to have weak influence on the kinematic properties, like the cell sizes $L/d$ and $W/d$ in the plane shear, the annular chute flow and the vertical chute, or even the rotating drum aspect ratio $D/d$ at given $Q^*$. The roughness of the boundaries have an influence on the boundary layers thickness. In the chute flow, the shear band width increases with the diameter of the beads glued on the wall. In the inclined plane, the material that covers the plane modifies the effective friction ($h_{\rm stop}$ and $h_{\rm start}$). If small beads of size $d$ are flowing on large glued ones, part of them remain blocked in the large holes thus creating an apparent roughness of size $d$. If on the contrary large beads are flowing on small glued ones, the effective friction is strongly reduced. In surface flows, the lateral boundaries induce a further friction proportional to the rolling height to the cell width ratio $h/W$. In these three cases, the wall influence can be encoded into effective macroscopic quantities, that can be measured. An important open problem for future studies is the transition between the different geometries. In principle, the linear shear should be recovered in the annular shear cell, in the limit of a very large radius of curvature $R_i/d$. When does the transition from a localised shear to a linear profile occurs? Similarly, a surface flow should be recovered on an inclined plane when the tilt angle tends to the dynamical angle. Can one obtain such surface flows in numerical simulations? Each of the possible transitions gives rise to new questions (chute flow vs inclined plane; plane shear vs surface flow; etc).

\subsection{Quasi-static versus inertial regimes}

Now that the relevant dimensionless parameters are determined, we can study the different flow regimes, which classically are divided in a quasi-static, dense inertial and collisional regimes. The data collected in this paper allows us to discuss the existence of the three regimes and the transition from one to another in the different geometries. We first analyse the case of the plane shear configuration. Figure~\ref{FigPlaneShear}e displaying the effective friction coefficient as a function of the dimensionless parameter $I$ shows three regimes. In the limit of zero $I$ the system is rate independent, $\mu_{\rm eff}$ is constant. When $I$ increases, the inertia starts influencing the flow and the system becomes rate dependent. Eventually, for large value of $I$, the flow becomes dilute and collisional. The fact that $I$ controls the transitions means that one can evolve from one regime to another either by increasing the shear rate or decreasing the pressure.
The transition between the dense regime and the collisional regime is clearly identified in figure~\ref{FigPlaneShear}e with the slope discontinuity of $\mu_{\rm eff}(I)$. By contrast, no sharp transition is observed between quasi-static and dense inertial flows: the effective friction coefficient continuously decreases towards its quasi-static value when $I$ decreases. The same conclusion can be drawn when looking at the annular shear case (fig.~\ref{FigAnnularShear}b). In the rotating drum, one even observe the spatial coexistence of a quasi-static creep motion where $I$ is very small -- the exponential tail of the velocity profile -- and an inertial dense flow -- the linear part of the velocity profile (Fig.~\ref{FigSurfaceFlow}c). For some of the authors, this observation indicates that the intermediate dense flow regime has more in common with the quasi-static regime than with the collisional one.

In the following we successively discuss in more detail the quasi-static limit and the inertial dense regime.

\subsection{Quasi-static regime}

The limit $I \to 0$ of the quasi-static regime is easily achieved in configurations where the deformation is imposed, namely in the plane shear flow, in the annular shear flow and in the vertical chute flow. In some cases, the flows can exhibit very intermittent behaviours also the mean velocity profiles remain smooth. Apart from the plane shear geometry, the velocity profiles observed in this regime exhibit localised shear bands close to wall, the width of the shear band being few particle diameters. This is indeed observed in the circular shear cell (Fig.~\ref{FigAnnularShear}c) and the chute flow (Fig.~\ref{FigChuteFlow}b). In this respect the exponential tail in the heap and drum geometry could be seen as a particular case of localisation, where the flowing layer plays the role of a driving wall (Fig.~\ref{FigSurfaceFlow}c et~\ref{FigSurfaceFlow}f). On the contrary, for some reason that remains to be explained, the simulations of the plane shear cell shows a velocity profile which is not localised but linear (Fig.~\ref{FigPlaneShear}b). It is not clear if this unexpected behaviour is related to the fact that the plane shear is the only geometry where the stresses are strictly uniform in the cell. It is also interesting to note that recent experiments of granular flow in the annular shear configuration~\cite{FV03} indicates that the shear band thickness can dramatically increase when changing the bottom boundary conditions. Altogether, the ingredients underlying the existence of localisation remain unclear and need further investigations.

 In the quasi-static limit, the velocity profiles (localised or not) are independent of the imposed shear rate. Shearing twice as fast gives the same velocity profile multiply by two (see for example Fig.~\ref{FigChuteFlow}e). However, one has to recall that the macroscopic time scale $T_{\gamma}$ associated to the shear rate is not the only time scale. The other typical time scale associated to the confining pressure $T_P$ (eq.~\ref{TP}) is not zero even in the quasi-static limit. As a result, granular flows in the quasi-static regime are not time invariant. Shearing the material twice as fast does not give the same movie played twice as fast: the typical time of the rapid events associated to one particle passing over another is not divided by two. This remark is of importance when analysing the velocity fluctuations $<\delta V^2>$, in both the plane shear (fig.~\ref{FigPlaneShear}g) or the annular shear (fig.~\ref{FigAnnularShear}h). In both cases the velocity fluctuations $\sqrt{<\delta V^2>}$ do not simply scale with $\dot \gamma d$. In the plane shear $\sqrt{<\delta V^2>}/d\dot \gamma $ varies like $I^{-1/2}$. In the annular shear $\sqrt{<\delta V^2>}/d\dot \gamma $ measure at the free surface scales with $I^{-2/3}$, $I$ being equal in this case to $\dot \gamma \sqrt{d/g}$. 

This scaling can be compared with prediction arising from the naive picture given in figure \ref{TimeScale}. Based on the idea of a process made of a succession of rapid events occurring on a time scale $T_P$, one can derive the mean velocity fluctuations in the limit $T_P \ll T_{\gamma}$. In this limit, $\delta V^2$ is of order $P/\rho$ during the rapid events, and is negligible in between. One can then write that the averaged velocity fluctuations is equal to $<\delta V^2>\simeq \frac{T_{P}~P/\rho }{T_P+T_{\gamma}}$. Using the expressions of $T_P$ and $T_{\gamma}$ eqs.~(\ref{Tgamma}) and (\ref{TP}), we then find that $\sqrt{<\delta V^2>}/d\dot \gamma \simeq I^{-1/2}$. This simple argument could then explain the scaling observed in the plane shear, but not the -2/3 of the annular shear. Nevertheless, this analogy suggests that the study of the whole velocity distribution in the quasi-static regime could give more information about the underlying mechanism than the simple rms fluctuations.

\subsection{Dense inertial regime}

When the parameter $I$ increases above $10^{-2}$, the effective friction coefficient is no longer constant but increases with $I$, indicating a shear rate dependent regime. The shear rate dependence is observed in the simulations of the plane shear cell in figure~\ref{FigPlaneShear}e, in the annular shear figure~\ref{FigAnnularShear}b, in the inclined plane in figure~\ref{FigInclinedPlane}e. However the plane shear geometry is the simplest configuration as the velocity profiles are linear. It is then tempting to conclude that the plane shear plays the role of a rheometer and that the relation $\mu_{\rm eff}(I)$ provides the local constitutive law for dense granular flow. One could then legitimately wonder if a simple local rheology stipulating that everywhere in the flow, stresses are related to shear rate through the relation $\tau/P=\mu(\dot \gamma d/\sqrt{P/\rho})$, where $\mu(I)$ has the shape of figure~\ref{FigPlaneShear}e, can describe all geometries of dense granular flows. In the following section we discuss the implication of this assumption in regards to the data collected in this paper, before discussing the necessity to consider a non local rheology. 

\subsubsection{The local rheology assumption}
Hereafter, we will call local, a rheology for which stresses and shear rate at a given location in the flow are related trough a univoque relation, which from dimensional analysis can be written as: 
\begin{equation}
\tau/P=\mu(I), 
\end{equation}
$I$ being given by eq.~(\ref{defI}). As soon as $\tau/P$ depends on the rescaled shear rate at other locations or on any further field whose governing equation would need to be precised, we consider the rheology as non local. 

Under this assumption of a local rheology, one can predict velocity profiles in both plane shear and inclined plane configurations. In the following we test this hypothesis facing the data. In the plane shear case, the stress distribution is uniform:
\begin{equation}
P=cte ,\quad \tau/P=cte.
\end{equation}
Accordingly, the parameter $I$ has to be constant across the cell, equal everywhere to $\mu^{-1}(\tau/P)$. The shear rate $\dot\gamma$ is then also uniform and the predicted velocity profile is linear. This is in agreement with the measurement in figure~\ref{FigPlaneShear}b for moderate $I$, before the flow becomes collisional. The measurement of the effective coefficient at the wall $\tau_w/P_w$ then coincides exactly  with the rheological law $\mu(I)$. 

In the case of surface flows, the stress distribution is the
following:
\begin{equation}
P=\rho g (h-y)\cos(\theta), \quad \tau/P=\tan(\theta)
\end{equation}
The shear rate $\dot\gamma$ is then selected by the relationship
\begin{equation}
I=\frac{\dot\gamma(y) d}{\sqrt{P(y)/\rho}}=\mu^{-1}(\tan(\theta))
\label{Itheolocal}
\end{equation}
Integrating $\dot\gamma$ in the above relation leads to a profile going like the depth to the power 3/2, the Bagnold like profile. 
\begin{equation}
\frac{V(y)}{\sqrt{gd}}=A(\theta)\frac{\left(h^{3/2}-(h-y)^{3/2}\right)}{d^{3/2}}
\label{bagnold}
\end{equation}
with
\begin{equation}
A(\theta)=\frac{2}{3} I(\theta) \sqrt{\cos(\theta)}.
\label{relationIA}
\end{equation}

It is worth noting that the Bagnold profile does not result from collisional arguments, but simply relies on dimensional reasoning. We can now compare the prediction of the local rheology with experimental measurements. We have seen that in the experiments, the depth averaged velocity $<V>$, the thickness $h$ and the inclination $\theta$ are related trough the scaling~\ref{frhstop}. The prediction of the local rheology for depth average velocity are obtained by integrating relation~\ref{bagnold} over the flow depth. One get the following relation between $<V>$, $h$, and $\theta$ : 

\begin{equation}
\frac{<V>}{\sqrt{gh}}=\frac{3}{5}\frac{h}{d} A(\theta)
\label{scalinglocal}
\end{equation}

The predicted scaling is then not fully compatible with the observed one as it does not predict the coefficient $\alpha$ of equation~\ref{frhstop}. However, it is compatible with the case of glass beads for which $\alpha=0$. In this case, equation~\ref{scalinglocal} together with~\ref{frhstop} implies that the Bagnold constant $A(\theta)$ has to be related to $h_{\rm stop}$ :

\begin{equation}
A(\theta)= \frac{5}{3} \beta \frac{d}{h_{\rm stop}(\theta)}
\label{Ahstop}
\end{equation}

\begin{figure}[h!] \bc
\includegraphics{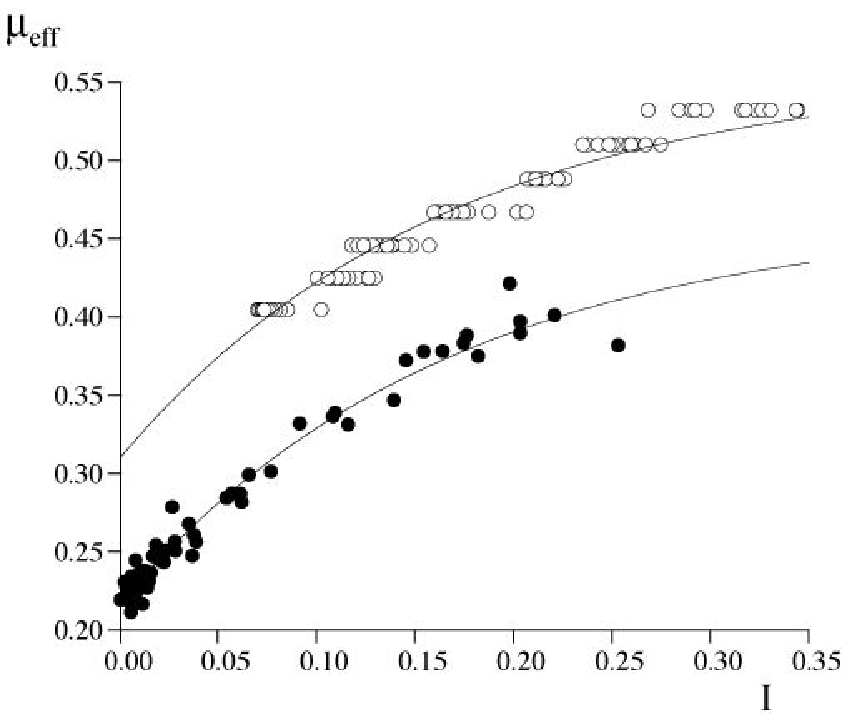}
\caption{Comparison of the effective friction function of $I$ in the plane shear ({\Large $\bullet$}) and the inclined plane ({\Large $\circ$}) configurations.}
\label{MusOfI}
\ec
\end{figure}

Finally, it is interesting to note that, under the assumption of the local rheology, the inclined plane configuration could be used also as a rheometer. The function $\mu(I)$ can indeed be measured as follows. By imposing the inclination, the experimentalist fixes the friction coefficient $\mu=\tan\theta$ and measures the corresponding parameter $I$. From equations (\ref{relationIA}) and (\ref{scalinglocal}) one find that $I$ should be related to the depth averaged velocity and thickness through the relation:

\begin{equation}
<I>=\frac{5}{2} \; \frac{<V>d}{h\sqrt{gh\cos{\theta}}}.
\label{IInclinedPlane}
\end{equation}

This can be tested from data of figure~\ref{FigInclinedPlane}. Figure~\ref{MusOfI} shows that the data for glass beads collapse relatively well when $\tan\theta$ is plotted as a function of $I$ given by equation~(\ref{IInclinedPlane}). On the same graph we have reported the $\mu(I)$ measured in the 2D simulation of the plane shear. Interestingly, the shape and the range of $I$ are similar. However, the same is not true for the other materials. As the coefficient $\alpha$ of equation~\ref{frhstop} is not zero for sand or 2D disks, the effective friction coefficient is not a function of $I$ only but depends also on $h/d$ (eq.~\ref{mueffinclinedplane}).

As a conclusion, the local rheology assumption captures some of the basic features observed in both the plane shear and the inclined plane geometries, in particular the scaling of the averaged quantities. However, for surface flows in the rotating drum and down the heap, whereas the stress distribution is the same as in the inclined plane, the observed velocity profiles in the flowing layer are linear and not Bagnold. In particular, the shear rate does not vanish at the free surface~\cite{Ra03}. Let us emphasise that this violation of the Bagnold profile is not to be attributed to the specific shape of $\mu(I)$. This will force us to relax the local rheology hypothesis. Let us postpone this to section~\ref{localreho} and introduce first a more general framework of analysis.

\subsubsection{Prandtl mixing length approach}
An alternative description can be provided by generalising Bagnold shear stress $\tau=\rho d^2 \dot \gamma^2$, as suggested by Orpe and Khakhar~\cite{OK01} and Ertas and Halsey~\cite{EH02}. The approach consists in introducing a correlation length scale $l$ instead of $d$, $l$ being related to the size of some clusters in the flow. This approach is reminiscent of Prandtl closure for turbulence flows where a turbulent viscosity is introduced equal to $\rho l^2 \dot \gamma$, $l$ being interpreted as the size of the large eddies. For our granular case, the shear stress is then written as: 

\begin{equation}
\tau=\rho l^2 \dot\gamma^2.
\label{prandtl}
\end{equation}

The case discussed in the previous section of a local rheology described by a friction $\mu(I)$ is a particular case of the mixing length description. Indeed, if $\tau/P=\mu(I)$ and $P=\rho d^2\dot\gamma^2/I^2$ by definition of $I$, one obtains:
\begin{equation}
\tau=\rho \frac{\mu(I) d^2}{I^2} \dot \gamma^{2}
\end{equation}
that is, a correlation length function of the {\it local} properties of the flow: $l(I)=\sqrt{\mu(I)} d/I$. 

Assuming a local dependence of the correlation length is then strictly equivalent to the local rheology case described before. However, it gives an insight to the results observed experimentally in the inclined plane. This is shown by first deriving the velocity profiles predicted by equation~\ref{prandtl}. One recovers the Bagnold profiles with a function $A(\theta)$ expressed in term of $l$ as follows : 
\begin{equation}
A(\theta)=\frac{2}{3} \sqrt{\sin \theta}\frac{d}{l}.
\label{Adetheta}
\end{equation}
We have seen that in order to be compatible with the scaling experimentally observed, the function $A(\theta)$ should be related to $h_{\rm stop}$ (eq.~\ref{Ahstop}). This means that the correlation length is related to the function $h_{\rm stop}$:
\begin{equation}
l(\theta)=\frac{2}{5} \frac{\sqrt{\sin(\theta)}}{\beta} h_{\rm stop}(\theta)
\end{equation}
The function $h_{\rm stop}$ actually measures the correlation length $l$, i.e. the characteristic size of coherent motions. This gives an interpretation to the existence of the flow threshold: no flow is possible when the thickness becomes less than few times the correlation length. Altogether, the local rheology assumption is equivalent to the Prandtl mixing length description with $l$ depending on $I$ only, which furthermore relates the inclined plane rheology to the deposit thickness $h_{\rm stop}$.

\subsubsection{Towards a non local rheology}
\label{localreho}
\begin{figure}[h!] \bc
\includegraphics{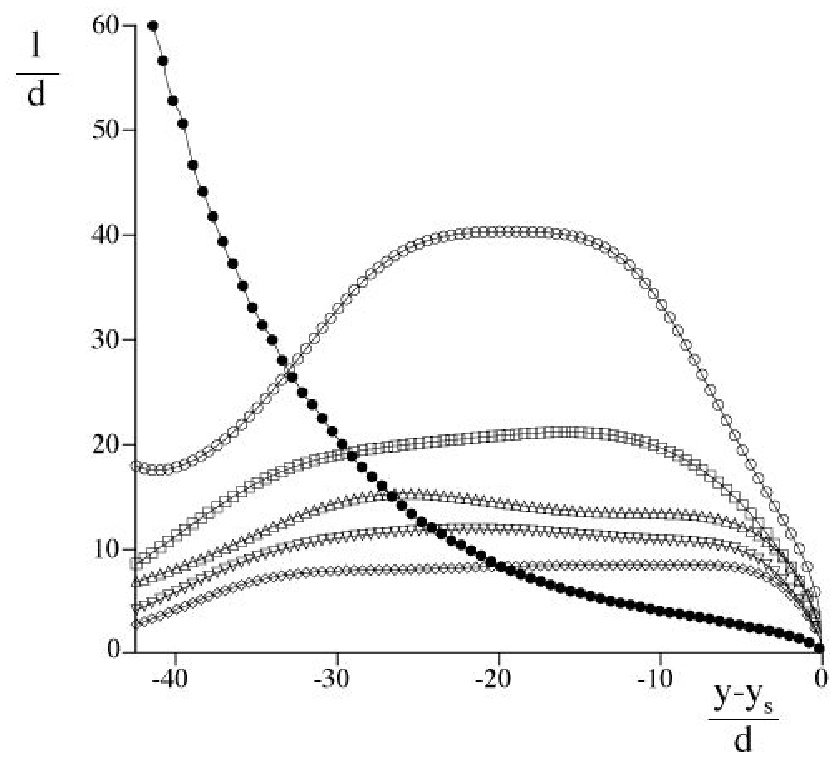}
\caption{Correlation length $l(y)$ obtained from inclined plane simulations IP4 for $\theta=15^\circ$ ({\Large $\circ$}), $\theta=16^\circ$ ($\square$), $\theta=17^\circ$ ($\triangle$), $\theta=18^\circ$ ($\triangledown$), $\theta=21^\circ$ ($\diamond$), and from heap case experiment SF4 ({\Large $\bullet$}).}
\label{ElOfZ}
\ec
\end{figure}

In the previous paragraphs, we have checked that the local rheology assumption is compatible with the average flow properties. We wish now to check its validity facing the velocity profiles. To do so, let us define a local correlation length $l(y)$ that a priori depends on the position $y$ :
\begin{equation}
l(y)^2=\frac{\tau(y)}{\rho\dot\gamma(y)^2}. 
 \end{equation}
If the local rheology is valid, $l(y)$ should be constant across the flowing layer. The numerical simulations or experimental measurements of the velocity profile provide $\dot\gamma(y)$ and the momentum balance gives $\tau(y)$, so that one can compute $l(y)$. Figure~\ref{ElOfZ} displays the $l(y)$ profiles in the inclined plane configuration for different inclinations $\theta$. For high inclinations and far from the bottom or the top, a plateau is observed, consistent with a constant $l$ predicted by the local rheology. The correlation length in the bulk decreases with $\theta$ just like $h_{\rm stop}$. However, the correlation length decreases at the top and bottom of the profile. These results enlighten the existence of some boundary layers which could be interpreted as regions where the grains feel the boundaries and hence experience different correlations with their neighbours. When decreasing the inclination i.e. getting closer to the flow threshold, the plateau disappears as shown in figure \ref{ElOfZ}. The correlation length becomes of the order of the thickness, meaning that everywhere the grains feel the boundaries.

Using the plateau value of $l$ obtained in the bulk of the flow, we can reconstruct the Bagnold profile according to eqs.~(\ref{bagnold}) and (\ref{Adetheta}). These profiles are plotted on figure~\ref{FigInclinedPlane}i. They emphasise the existence of deviations close to the bottom and to the free surface, which become more and more important and invade the whole layer at low inclinations. 

Performing the same analysis for the flow on a heap, no plateau is observed (circles in fig.~\ref{ElOfZ}). The correlation length $l(y)$ vanishes at the free surface, increases up to the transition toward the static phase where it diverges. This behaviour, drastically  different from the expected constant correlation length, definitely calls for the introduction of a non-local rheology. 

From a general point of view, one could introduce a non-local dependence of $\tau/P$ on $I$ or introduce further fields such as $\Phi$ together with their constitutive relations. A simple way of doing this in the Prandtl mixing length framework is to choose a correlation length which depends on the distance to the free surface \cite{KOAO01}. This is what is classically proposed to describe turbulent boundary layer. The observed Log velocity profile is recovered by assuming $l$ to be proportional to the distance to the wall. To be more specific, choosing
\begin{equation}
l^2=B(\theta) d (h-y)
\end{equation}
leads to a linear velocity profile with a shear rate
\begin{equation}
\dot\gamma=\sqrt{\frac{g\sin(\theta)}{B(\theta)}}
\end{equation}

In conclusion, the description in term of a correlation length theory, although very crude, allows to describe in the same formalism the transition from the flow on a rough plane to that on a heap. It strongly suggests to focus on correlations in granular flows to better understand what control coherent motions and to characterise the non local mechanisms. As a matter of fact, structures like arches \cite{CPMDBGCR01,P02}, dense correlated grains clusters \cite{BDL01b,B01,K99,DTdM03,RR03} and non-local dissipation due to multiple collisions \cite{AD01} have been evidenced experimentally or numerically . They have motivated several non-local models \cite{MLT99,PFL01,RC02,Ra03}. At the present time, none of these models have succeeded in rendering the kinematic properties of dense granular flows throughout the various geometries. This remains a challenging problem, which calls for further efforts in identifying the origin of non-locality.

\section{ Conclusion}

By collecting data coming from different groups, both in experiments and simulations, in different geometries, we have been able to capture important characteristics of granular flows. The relevant time scales and length scales have been identified, the transition between flow regimes has been clarified and some important ingredients such as the necessity of considering non local rheology have been discussed. However, we are far from the end of the story. When trying to compare and extract common physical mechanisms among the different granular flows, many open problems have emerged that will provide work for the future. If a single conclusion has to be formulated about this collective work, it would be that, at the present time, with our knowledge of granular flows and the amount of data available, one can no longer consider a single geometry as a test for constitutive law but should consider the different geometries.

\begin{acknowledgement}
{\bf Acknowledgments\\} This work is the result of a common work of the Groupement de Recherche sur les Milieux Divis\'es (GDR Midi 2181, CNRS), which gathers the French laboratories involved in granular media. The data have been collected by Bruno Andreotti, François Chevoir, Olivier Dauchot, Olivier Pouliquen and Patrick Richard. It would not have been possible without the administrative help of Jeanne Pullino, Nelly Sammut and Frédérique Oger. 
\end{acknowledgement}

\appendix
\section{Notation}

\begin{itemize}

 \item[$\bullet$] {\bf Geometrical parameters}

 \item[] $x$: flow direction
 \item[] $y$: direction transverse to the flow
 \item[] $\theta$: flow inclination

 \item[] $L$: distance between the walls in the confined flow cases
 \item[] $W$: distance in the invariant spanwise direction
 \item[] $R_{i,o}$: inner, resp. outer cylinder radii of the annular shear cell.

 \item[] $R$: drum radius
 \item[] $D=2R$: drum diameter

 \item[] $\Omega$: angular velocity of the rotating drum or of the annular shear cell inner cylinder
 \item[] $\Gamma$: torque applied to the inner cylinder of the annular shear cell
 \item[] $Q$: flow rate

 \item[] $V_{\rm w}$: wall velocity
 \item[] $\dot\gamma_{\rm w}$: characteristic shear rate $V_{\rm w}/L$ in plane shear, $V_{\rm w}/d$ in annular shear
 \item[] $P_{\rm w}$: wall pressure
 \item[] $\tau_{\rm w}$: wall shear stress
 \item[]
 \item[$\bullet$] {\bf Microscopic parameters}

 \item[] $d$: particle diameter
 \item[] $d_{\rm w}$: diameter of the particles eventually glued on the walls
 \item[] $e$: restitution coefficient
\item[] $\mu_{\rm p}$: inter-particles friction coefficient
 \item[] $ \mu_{\rm w}$: particle-wall friction coefficient
 \item[] $g$: gravity
 \item[]
 \item[$\bullet$] {\bf Measured quantities}

 \item[] $y_s$: free surface $y$-coordinate
 \item[] $y_b$: flow-no flow interface $y$-coordinate
 \item[] $h=y_s-y_b$: flow thickness

 \item[] $\theta_{\rm start}$; $\theta_{\rm stop}$: limiting angles at which the flow starts (resp. stops).
 \item[] $h_{\rm start}$; $h_{\rm stop}$: limiting thickness at a given angle at which the inclined plane flow starts (resp. stops)

 \item[] $V(y)$: time averaged velocity profile
 \item[] $\delta V^2(y)$: time averaged squared velocity fluctuations
 \item[] $\dot\gamma=\frac{dV}{dy}$: shear rate
 \item[] $\Phi(y)$: volume fraction profile
 \item[] $\rho$: flow mass density
 \item[] $<.>$: averaging operator over the flow thickness
 \item[] $\sigma$: stress tensor
 \item[] $\tau=\sigma_{xy}$: shear stress
 \item[] $P=\sigma_{yy}$: normal stress
 \item[] $\mu_{\rm eff}=\tau/P$: effective friction coefficient

 \item[]
 \item[$\bullet$] {\bf Dimensionless quantities}

 \item[] $I=\frac{\dot\gamma d}{\sqrt{P/ \rho}}$: dimensionless shear rate
 \item[] $Fr=\frac{<V>}{\sqrt{gh}}$: Froude number
 \item[] $Q*=\frac{Q}{d \sqrt{gd}}$

\end{itemize}

\bibliographystyle{unsrt}
\bibliography{ArtGDR3}
\end{document}